\documentclass[varenna]{cimento}

\usepackage{latexsym}
\usepackage{amstext}
\usepackage{amsxtra}
\usepackage{graphicx}
\usepackage{amssymb}

\newcommand{\bs}{\boldsymbol}

\newcommand{\DctwoD}{D_{\rm c}}

\newcommand{\NctwoDid}{N_{\rm c}^{({\rm id})}}

\newcommand{\NctwoDmf}{N_{\rm c}^{\rm (mf)}}

\newcommand{\nthree}{n_3}

\newcommand{\gthreeD}{g^{\rm (3d)}}

\newcommand{\kB}{k_{\rm B}}

\begin{document}
\title{Two-dimensional Bose fluids: \\ An atomic physics perspective}
\author{Zoran Hadzibabic}
\institute{Cavendish Laboratory, University of Cambridge \\ JJ Thomson Avenue, Cambridge CB3 0HE, United Kingdom}
\author{Jean Dalibard}
\institute{Laboratoire Kastler Brossel, CNRS, UPMC, \'Ecole normale sup\'erieure \\ 24 rue Lhomond, 75005 Paris, France}

\shortauthor{Zoran Hadzibabic \atque Jean Dalibard}

\maketitle

\setcounter{secnumdepth}{2}
\setcounter{tocdepth}{1}

\begin{abstract}
We give in this lecture an introduction to the physics of two-dimensional (2d) Bose gases. We first discuss the properties of
uniform, infinite 2d Bose fluids at non-zero-temperature $T$. We
explain why thermal fluctuations are strong enough to destroy the
fully ordered state associated with Bose--Einstein condensation, but
are not strong enough to suppress superfluidity in an interacting
system at low $T$. We present the basics of the
Berezinskii--Kosterlitz--Thouless theory, which provides the general
framework for understanding 2d superfluidity. We then turn to
experimentally relevant finite-size systems, in which the presence
of residual ``quasi-long-range" order at low temperatures leads to
an interesting interplay between superfluidity and condensation.
Finally we summarize the recent progress in theoretical
understanding and experimental investigation of ultra-cold atomic
gases confined to a quasi 2d geometry. 
\end{abstract}

\newpage


\section{Introduction}

The properties of phase transitions and the types of order present in the low-tempera\-ture states of matter are fundamentally dependent on the dimensionality of physical systems~\cite{Peierls:Surprises}. Generally, highly ordered states are more robust in higher dimensions, while thermal and quantum fluctuations, which favor disordered states, play a more important role in lower dimensions.

The case of a two-dimensional (2d) Bose fluid is particularly fascinating because of its ``marginal" behavior. In an infinite uniform 2d fluid thermal fluctuations at any non-zero temperature are strong enough to destroy the fully ordered state associated with Bose--Einstein condensation, but are not strong enough to suppress superfluidity in an interacting system at low, but non-zero temperatures. Further, the presence of residual ``quasi-long-range" order at low temperatures leads to an interesting interplay between superfluidity and condensation in all experimentally relevant finite-size systems.

This behavior is characteristic of a wide range of physical systems which share some generic properties such as dimensionality, form of interactions, and Hamiltonian symmetries. These include liquid helium films~\cite{Bishop:1978}, spin-polarized hydrogen~\cite{Safonov:1998a}, Coulomb gases~\cite{Minnhagen:1987}, ultra-cold atomic gases~\cite{Posazhennikova:2006,Bloch:2008}, exciton~\cite{Snoke:2002, Butov:2004} and polariton~\cite{Kasprzak:2006, Amo:2009} systems. Moreover, the Berezinskii--Kosterlitz--Thouless theory which provides the general framework for understanding 2d superfluidity is also applicable to a range of physical phenomena in discrete systems, such as ordering of spins on a 2d lattice~\cite{Kosterlitz:1974} and melting of 2d crystals~\cite{Nelson:1979, Strandburg:1988}.

In this paper we give an introduction to the physics of 2d Bose fluids from an atomic physics perspective. Our goal is to summarize the recent progress in theoretical understanding and experimental investigation of ultra-cold atomic gases confined to 2d geometry, and we also hope to provide a useful introduction to these systems for researchers working on related topics in other fields of physics.


\subsection{Absence of true long-range order in 2d}

The most familiar phase transitions in three spatial dimensions (3d), such as freezing of water, ferromagnetic ordering in spin systems, or Bose--Einstein condensation (BEC), are all associated with emergence of true long-range order (LRO) below some non-zero critical temperature. Such order is embedded in a spatially uniform order parameter, e.g. magnetization in a ferromagnet or the macroscopic wavefunction $\psi$ describing a BEC. Further, in all the above cases emergence of true LRO corresponds to spontaneous breaking of some continuous symmetry of the Hamiltonian. In case of crystallization (freezing), translational symmetry is spontaneously broken. For Heisenberg spins on a (fixed) lattice, the Hamiltonian has a continuous spin rotational symmetry which is spontaneously broken in the ferromagnetic state. In case of BEC, the phase of $\psi$ is arbitrarily spontaneously chosen at the transition. As we will discuss in more detail later, under certain conditions this makes a Bose gas formally equivalent to a system of two-component spins on a lattice, the so-called XY model.

Already in 1934 Peierls pointed out that the possibility for a physical system to exhibit true LRO can crucially depend on its dimensionality~\cite{Peierls:1934,Peierls:1935}. Peierls considered a 2d crystal at finite temperature $T$, and studied the effects of thermal vibrations of atoms around their equilibrium positions in the lattice (i.e. phonons). He found that the uncertainty in the relative position of two atoms \textit{diverges} with the distance between their equilibrium positions:
\begin{equation}
\label{Peierls}
\left\langle \left( \bs u (\bs r) - \bs u (0) \right)^2 \right\rangle \propto
T \ln\left(\frac{r}{a}\right)
\end{equation}
for $r= |\bs r| \gg a$, where $\bs u \left(\bs r \right)$ is the atom displacement from its equilibrium position $\bs r$, $a$ is the lattice spacing, and $\left\langle ... \right\rangle$ denotes a thermal average. In contrast, the corresponding result in 3d is finite, and small compared to $a$ if $T$ is below the melting temperature. The result in eq.~(\ref{Peierls}) is in direct contradiction with the starting hypothesis of long-range crystalline order, since it implies that, based on the positions of atoms in one part of the system, we cannot predict with any certainty the positions of atoms at large distances.

The Peierls result (\ref{Peierls}) is a simple example of the absence of spontaneous symmetry breaking at non-zero $T$ in 2d. The absence of LRO in low-dimensional systems was later more generally and formally studied by Bogoliubov~\cite{Bogoliubov:1960}, Hohenberg~\cite{Hohenberg:1967}, and Mermin and Wagner~\cite{Mermin:1966}. The general statement is that LRO is impossible in the thermodynamic limit at any non-zero $T$ in all 1d and 2d systems with short-ranged interactions and a continuous Hamiltonian symmetry. This is now most commonly known as the Mermin--Wagner theorem. In all such systems Hamiltonian symmetry is always restored by low-energy long-wavelength thermal fluctuations, the so-called Goldstone modes. In the case of an interacting Bose gas Goldstone modes are phonons, while in the case of the XY model on a lattice they are spin-waves. As a direct consequence of the functional form of the density of states in low dimensionality, such modes always have a diverging infrared contribution and destroy LRO. It is however important to stress that the absence of true LRO does not preclude the possibility of any phase transitions in 2d systems, just the symmetry breaking ones. As we will see later, a phase transition associated with the apparition of a \emph{topological order} does take place in a uniform 2d gas.


\subsection{Outline of the paper}

In Section~\ref{sec: uniform 2d} we will explicitly see that Bose--Einstein condensation is impossible in both the ideal and a repulsively interacting infinite uniform 2d Bose gas. The absence of true LRO in these systems is seen in the fact that the first order correlation function:
\begin{equation}
\label{define g1}
g_1(\bs r) \equiv \left\langle \hat{\Psi}^\dag(\bs r) \hat{\Psi}(0) \right\rangle \; ,
\end{equation}
where $\hat{\Psi}(\bs r)$ is the annihilation operator for a particle at position $\bs r$, always tends to zero for $r \rightarrow \infty$. Note that these two statements are equivalent under the Penrose-Onsager definition of the condensate density~\cite{Penrose:1956}:
\begin{equation}
\label{define PO}
n_0  \equiv \lim_{r \rightarrow \infty} g_1(\bs r) \; .
\end{equation}

However, the weak logarithmic divergence in eq.~(\ref{Peierls}) suggests that the destruction of LRO is only marginal in 2d.
The consequence of this weak divergence is that an interacting Bose gas at low $T$ exhibits ``quasi-long-range order", corresponding to $g_1(\bs r)$ which decays only algebraically with distance.

This low $T$ state is also superfluid, and the phase transition between the superfluid and the high $T$ normal state is described by the Berezinskii--Kosterlitz--Thouless (BKT) theory~\cite{Berezinskii:1971, Kosterlitz:1973}, which we discuss in Section~\ref{uniform BKT}. Further, the slow decay of $g_1(\bs r)$ at low $T$ has important consequences for condensation and symmetry breaking in the experimentally relevant finite-size systems.  We address this issue in Sections~\ref{sec:2d gas in a box} and~\ref{sec:2d gas in a trap}, first for a finite box potential, and then for the experimentally most pertinent case of a harmonically trapped gas.

In Sections~\ref{sec:experimental realizations} and~\ref{sec:experimental probes} we introduce the experimental methods of atomic physics used in the current studies of 2d Bose gases. In \S~\ref{sec:experimental realizations} we give an overview of experimental systems in which (quasi-)2d Bose gases have been realized, and in \S~~\ref{sec:experimental probes} we discuss the experimental probes of coherence in these systems. We conclude by outlining some research directions which are likely to be of interest in the near future in Section~\ref{sec:conclusions}.


\section{The infinite uniform 2d Bose gas at low temperature}
\label{sec: uniform 2d}

In this section we discuss thermal fluctuations and the absence of Bose--Einstein condensation in an infinite uniform 2d gas. In \S~\ref{ideal 2d Bose gas} we consider the ideal gas, in which no phase transition occurs. In this case the first order correlation function $g_1(\bs r)$ gradually changes from a gaussian function in the high-temperature, non-degenerate regime, to an exponentially decaying function in the degenerate regime.

In the repulsively interacting 2d Bose gas the BKT phase transition to a superfluid state occurs at a non-zero critical temperature, but the conclusion that BEC transition does not occur remains true. We first introduce the description of interactions in an atomic 2d gas in \S~\ref{sec:tilde g}. In \S~\ref{sec:density fluctuations} we qualitatively discuss why interactions lead to a strong suppression of density fluctuations in a degenerate gas, so that the low-energy long-wavelength excitations (phonons) in this system are almost purely phase fluctuations. The Bogoliubov analysis presented in \S~\ref{subsubsection:bogoliubov} provides a more quantitative justification for this conclusion and also indicates why we expect an interacting 2d gas to be superfluid at very low $T$. As we show in \S~\ref{sec:algebraic decay}, the long-wavelength phonons still lead to a vanishing $g_1(\bs r)$ at $r \rightarrow \infty$. Therefore, in accordance with the Mermin--Wagner theorem, these ``soft" Goldstone modes still destroy the long-range order and restore the Hamiltonian symmetry. However, the decay of $g_1$ at large distances is only algebraic with $r$ at very low $T$. This low temperature, superfluid state is said to exhibit ``quasi-long-range order". The BKT phase transition from the superfluid state with algebraic correlations to the normal state with exponentially decaying correlations is discussed in the following Section~\ref{uniform BKT}.


\subsection{The ideal 2d Bose gas}
\label{ideal 2d Bose gas}

The absence of Bose--Einstein condensation in the ideal 2d Bose gas can straightforwardly be seen by following the standard Einstein's argument which associates condensation with the saturation of the excited single-particle states at some non-zero temperature.

In 2d, the density of states for spinless bosons $m L^2 / (2\pi \hbar^2)$ is constant, where $m$ is the particle mass and $L \rightarrow \infty$ is the linear size of the system. Assuming no condensation, the total number of particles is then:

\begin{equation}
N = \frac{m L^2}{2\pi \hbar^2} \int_0^\infty \frac{d \varepsilon}{e^{\beta(\varepsilon - \mu)} -1}\ ,
\end{equation}
where $\beta = 1/(k_{\rm B}T)$, and $\mu \le 0$ is the chemical potential. Equivalently, the phase-space density $D$ is given by:
\begin{equation}
D\equiv n\lambda^2 = \int_0^\infty \frac{dx}{\frac{1}{Z} e^x - 1} = -\ln(1-Z)\ ,
\end{equation}
where $n=N/L^2$ is the 2d number density, $\lambda = h/\sqrt{2\pi m k_{\rm B}T}$ is the thermal wavelength, and $Z=e^{\beta \mu}$ is the fugacity.

In 3d, the signature of BEC is that the analogous relationship between the phase space density and fugacity has no solutions for $Z$ when the phase space density is larger than the critical value $\nthree \lambda^3 \approx 2.612$, where $\nthree$ is the 3d number density. Below the condensation temperature, chemical potential is fixed at $\mu =0$ and the phase space density of particles in the excited states is saturated at $\approx 2.612$. However, we see that in 2d a valid solution:
\begin{equation}
e^{\beta \mu} = 1 - e^{- n\lambda^2}
\label{no saturation}
\end{equation}
always exists. In other words, for any non-infinite phase space density there exists a negative value of $\mu$ which allows normalization of the thermal distribution of particles in the excited states to the total number of particles in the system $N$. This shows that BEC does not occur in the ideal infinite uniform 2d Bose gas.

We next look at the first order correlation function $g_1(r)$, which we can write as the Fourier transform of the momentum space distribution function $n_{\bs k}$:
\begin{equation}
g_1(r) = \frac{1}{(2\pi)^2}\int_0^\infty   n_{\bs k}\;e^{i \bs k \cdot \bs r}\;d^2 k\ ,
  \qquad \mbox{with}\quad
 n_{\bs k}=\frac{1}{e^{\beta(\epsilon_k-\mu)}-1}\ ,\quad \epsilon_k=\frac{\hbar^2k^2}{2m}\ .
\end{equation}
In absence of condensation $g_1(r)$ always vanishes at $r \rightarrow \infty$. However, it still shows qualitatively different behavior at high and low temperature:

\begin{itemize}
\item
In a non-degenerate gas, eq.~(\ref{no saturation}) gives $Z \approx n\lambda^2 \ll 1$. In this regime $|\mu| \gg k_{\rm B}T$ and all momentum states are weakly occupied:
\begin{equation}
n_{\bs k} \approx Z\, e^{-\beta \epsilon_k} \approx n \lambda^2 \, e^{-{k^2 \lambda^2}\,/\,{4 \pi}} \ll 1 \quad (\forall \bs k).
\end{equation}
In this case $g_1(r)$ is gaussian, showing only short-range correlations which decay on the length scale $\lambda/\sqrt{\pi}$:
\begin{equation}
\label{gaussian}
g_1(r) \approx n\, e^{-{\pi r^2}/{\lambda^2}} \; .
\end{equation}

\item
In a degenerate gas with $n\lambda^2 > 1$, from eq.~(\ref{no saturation}) we get $Z\approx 1$ and $\beta |\mu| \approx e^{- n\lambda^2} \ll 1$, so that $D = n \lambda^2 \approx \ln (k_{\rm B}T/|\mu|)$. In this regime the occupation of high energy states, with $\beta \epsilon_k \gg 1$, is still small and given by the Boltzmann factor:
\begin{equation}
n_{\bs k} \approx e^{-\beta \epsilon_k} = e^{-{k^2 \lambda^2}\,/\,{4 \pi}} \ll 1 \ , \ {\rm for} \: k^2 \gg 4\pi/\lambda^2 \; .
\end{equation}
However, the low energy states with $\beta \epsilon_k \ll 1$ are strongly occupied:
\begin{equation}
\label{lorentzian}
n_{\bs k} \approx \frac{k_{\rm B}T}{\epsilon_k + |\mu|} = \frac{4\pi}{\lambda^2}\, \frac{1}{k^2 + k_c^2} \gg 1 \ , \ {\rm for} \: k^2 \ll 4\pi/\lambda^2 \; ,
\end{equation}
where $k_c = \sqrt{2m |\mu|}/\hbar$. In this case $g_1(r)$ is bimodal. At short distances, up to $r \sim \lambda$, correlations are still gaussian as in eq.~(\ref{gaussian}). However, the Lorentzian form in eq.~(\ref{lorentzian}) corresponds to approximately~\footnote{More precisely the Fourier transform of the 2d Lorentzian distribution $1/(k^2+k_c^2)$ is defined for $\bs r\neq 0$ and is proportional to the Bessel function of imaginary argument $K_0(k_cr)$, whose asymptotic behaviour is $e^{-k_c r}/\sqrt r$.} exponential decay of $g_1(\bs r)$ at larger distances, $r \gg \lambda$:
\begin{equation}
g_1(r) \approx e^{-r/\ell} \ , \ {\rm with} \ \ell= k_c^{-1} \approx \lambda \, e^{n\lambda^2/2}/\sqrt{4\pi}.
\end{equation}
We can also estimate the partial phase space densities corresponding to the gaussian and the Lorentzian parts of the momentum distribution, and see that most particles accumulate in the Lorentzian part corresponding to low momentum states:
\begin{eqnarray}
D_{\rm G} &\approx& \frac{\lambda^2}{(2\pi)^2}\int_{\sqrt{4\pi/\lambda^2}}^\infty n_k \, d^2k \ \approx\ 1/e \ \ll \ D \qquad
\\
D_{\rm L} &\approx& \frac{\lambda^2}{(2\pi)^2}\int_0^{\sqrt{4\pi/\lambda^2}} n_k \; d^2k \ \approx\  D
\end{eqnarray}

\end{itemize}

We thus see that even though there is no phase transition in this system, the first order correlation function gradually changes from a gaussian with short range correlations in the non-degenerate regime, to an exponential in the degenerate regime. Further, for $n\lambda^2 >1$, the correlation length $\ell \propto e^{n\lambda^2/2}$ grows exponentially. Therefore, while $g_1(r)$ formally vanishes at $r \rightarrow \infty$ at any non-zero $T$, the length scale over which it decays is exponentially large in the deeply degenerate ideal Bose gas. This has important consequences for the finite-sized experimental systems, since for any arbitrarily large fixed system size $L$ there exists a low enough non-zero temperature $T$ for which $\ell > L$, and correlations span the entire system. This issue will be addressed in Sections \ref{sec:2d gas in a box} and \ref{sec:2d gas in a trap}.


\subsection{Interactions in a 2d Bose gas at low $T$}
\label{sec:tilde g}

The interaction between two atoms at positions $\bs r_i$ and $\bs r_j$ in a 3d Bose gas at low temperature is well characterized by the contact potential~\footnote{Strictly speaking, the delta function in 3d must be regularized by using for example the pseudo-potential~\cite{Huang:1987}. The extension of the notion of zero-range potential to the 2d case is discussed in~\cite{Olshanii:2002a} (see also~\cite{AlKhawaja:2002} for a discussion in terms of many-body $T$ matrix).}:
\begin{equation}
V(\bs r_i - \bs r_j) = \frac{4\pi\hbar^2}{m}a_s \, \delta^{(3\rm d)}(\bs r_i - \bs r_j)\ ,
\end{equation}
where $a_s$ is the 3d s-wave scattering length, and mechanical stability requires repulsive interactions $a_s>0$. In 2d the two-body scattering problem is in general  more complicated and the scattering amplitude is energy-dependent~\cite{Adhikari:1986}. We will address this issue in more detail in Section~\ref{subsection: 2d interactions}. However, in all experimentally relevant situations so far, the analysis of interactions is simplified by the fact that while the gas is kinematically 2d, the interactions can be described by 3d scattering. The condition for this simplification is that the thickness of the sample $\ell_0$ is much larger than the 3d scattering length $a_s$. In all current atomic experiments the ratio $\ell_0/a_s$ is larger than 30. In this case we can to a very good approximation write the interaction energy as:
\begin{equation}
E_{\rm int} = \frac{g}{2}\int  n^2(\bs r) \;d^2 r \: ,
\label{eq:interaction_energy}
\end{equation}
where $g$ is the energy-independent interaction strength and $n(\bs r)$ is the local density. Note that here and in the following we treat the density $n(\bs r)$ -- and later the phase $\theta(\bs r)$ -- as a classical function. This will notably simplify the mathematical aspects of our approach, while capturing all important physical consequences related for example to quasi-long range order and to the normal-superfluid transition.

In 3d, the interaction strength $g^{(3\rm d)} = (4\pi\hbar^2 / m) a_s$ explicitly depends on the scattering length. However, we see on dimensional grounds that in 2d we can write:
\begin{equation}
g = \frac{\hbar^2}{m} \tilde{g} \; ,
\end{equation}
so that $\tilde{g}$ is a dimensionless coupling constant. The 2d healing length, which gives the characteristic length scale corresponding to the interaction energy, is then given by:
\begin{equation}
\label{healing length}
\xi = \hbar / \sqrt{mgn} = 1/\sqrt{\tilde{g}n} \; .
\end{equation}

We can anticipate on dimensional grounds that $\tilde{g} \sim a_s/\ell_0$, corresponding to $n = \nthree \ell_0$. Specifically, if a gas is harmonically confined to 2d, say in the $x-y$ plane, we will see in Section~\ref{subsection: 2d interactions} that the expression for $\tilde{g}$ is:
\begin{equation}
\tilde{g} = \sqrt{8 \pi} \, \frac{a_s}{a_z} \; ,
\end{equation}
where in this case $\ell_0=a_z$ is the oscillator length along the kinematically frozen direction $z$.

We can qualitatively define the strongly interacting limit by the value of $\tilde{g}$ for which the interaction energy of $N$ particles, $E_{\rm int}$, reaches the kinetic energy $E_{\rm K}$ of $N$ non-interacting particles equally distributed over the lowest $N$ single-particle states. In this limit, we expect the many-body ground state to be strongly correlated. We can estimate the interaction energy with a mean-field approximation by setting $n(\bs r) = n$ in eq.~(\ref{eq:interaction_energy}), which corresponds to neglecting density fluctuations:
\begin{equation}
E_{\rm int} = \frac{\hbar^2}{2m} \tilde{g} N n \; .
\end{equation}
Using the 2d density of states, $m L^2 / (2\pi \hbar^2)$, the energy of the $N$-th excited single-particle state is $E_N = (2\pi \hbar^2 / m)n$, and so:
\begin{equation}
E_{\rm K} = \frac{1}{2}N E_N = \frac{\pi \hbar^2}{m} N n\; .
\end{equation}
The strongly interacting limit then corresponds to:
\begin{equation}
\label{strong g}
E_{\rm int} = E_{\rm K} \Rightarrow \tilde{g} = 2\pi\;.
\end{equation}
For comparison, the value of $\tilde{g}$ in the current experiments on atomic 2d Bose gases varies between $\sim 10^{-1}$~\cite{Hadzibabic:2006, Kruger:2007} and $\sim 10^{-2}$~\cite{Clade:2009}, while in the more strongly interacting $^4$He films~\cite{Bishop:1978} it is estimated to be of order $1$~\cite{Prokofev:2001}.

The fact that result (\ref{strong g}) is independent of the density of the gas $n$ is a natural consequence of the fact that $\tilde{g}$ is dimensionless. This is in contrast to the 3d case, where the relative importance of interactions is characterized by the dimensionless parameter $\nthree a_s^3$.

\subsection{Suppression of density fluctuations and the low-energy Hamiltonian}
\label{sec:density fluctuations}

At strictly $T=0$, a weakly interacting 2d Bose gas is condensed and described by a constant macroscopic wavefunction (a uniform order parameter) $\psi = \sqrt{n}e^{i\theta}$, where $n$ and $\theta$ are classical fields. At any finite $T$, both the amplitude and the phase of $\psi$ show thermal fluctuations. However, repulsive interactions will always lead to a reduction of density fluctuations in a low temperature gas. The interaction energy is $(g/2) \int  n^2(\bs r) \;d^2 r \;= (g/2) L^2 \langle n^2(\bs r) \rangle$, and so keeping the average density $n = \langle n(\bs r) \rangle$ fixed, we see that minimizing the interaction energy is equivalent to minimizing the density fluctuations:
\begin{equation}
(\Delta n)^2 = \langle n^2(\bs r) \rangle - n^2 = \left(g_2(0) - 1 \right)n^2 \; ,
\end{equation}
where $g_2(r) = \langle n(\bs r) n(0) \rangle/n^2$ is the \emph{normalized} second order (density-density) correlation function~\footnote{Note that following the conventions in the literature on different topics we normalize $g_2$ so that it is dimensionless, while $g_1$ has units of density.}. In the ideal Bose gas $g_2(0) = 2$, while if the density fluctuations are completely suppressed $g_2(0) = 1$.

We can estimate the minimum cost of density fluctuations by equating it with the increase in interaction energy from adding a single particle to the system:
\begin{equation}
\frac{\partial E_{\rm int}}{\partial N} = gn = \frac{\hbar^2}{m} \tilde{g} n \; ,
\end{equation}
where we have set $g_2(0)=1$, which is consistent with the density fluctuations being significantly suppressed. Comparing this with the thermal energy $k_{\rm B}T$, we get:
\begin{equation}
\frac{gn}{k_{\rm B}T} = \frac{\tilde{g}}{2\pi}D \; .
\end{equation}
This strongly suggests that at sufficiently low temperature, given by $D \gg 2\pi / \tilde{g}$, any density fluctuations must be very strongly suppressed. Numerical calculations~\cite{Prokofev:2001} show that they can be significantly suppressed already for $D \gg 1$, with the exact extent of the suppression depending on the strength of interactions $\tilde{g}$.

In the limit of strong suppression of density fluctuations, the interaction energy becomes just an additive constant ($(1/2) g n^2 L^2$) in the Hamiltonian, and the kinetic energy arises only from the variations of $\theta$, the phase of $\psi$. The system is then often described by an effective low-energy Hamiltonian:
\begin{equation}
\label{theta hamiltonian}
H_\theta = \frac{\hbar^2}{2m} n_s \int (\nabla \theta )^2\; d^2 r,
\end{equation}
where one heuristically replaces the total density $n$ with the (uniform) superfluid density $n_s \leq n$. This is physically motivated because one expects only the superfluid component to exhibit phase stiffness and to flow under an imposed variation of $\theta$, with local velocity of the superfluid given by $\bs v_s = (\hbar/m){\bs \nabla} \theta$. Also note that at $T=0$ superfluid density is equal to the total density, and at very low $T$ they are similar. In essence, renormalizing $n$ to the lower $n_s$ is an effective way of absorbing all the short distance physics, including any residual density fluctuations, and $H_\theta$ provides a good description of long-range physics, at distances $r \gg \xi, \lambda$.

The effective low-energy Hamiltonian $H_\theta$ is the continuous version of the Hamiltonian of the XY model of spins on a lattice. It can be used to derive the correct long-range algebraic decay of $g_1(r)$ in the low $T$ superfluid state (see \S~\ref{sec:algebraic decay}). However, it is important to stress some caveats:

(1) Even though at low temperature $n_s \approx n$, $H_\theta$ fundamentally cannot be the correct microscopic Hamiltonian. We can see this on very general grounds since the proper Hamiltonian is by definition a temperature-independent entity, while $H_\theta$ depends on the temperature through $n_s$. More accurately  $H_\theta$ represents the increase of the \emph{free energy} of the gas if one imposes the superfluid current $(\hbar/m){\bs \nabla} \theta$, for instance by setting the gas in slow rotation  (see the appendix in \cite{Bloch:2008} for more details).

(2) If and only if the density fluctuations are completely absent can the 2d Bose gas be formally mapped onto the XY model. This condition is essentially fulfilled for $D \gg 2\pi/\tilde{g}$, but it is not satisfied at the BKT critical point. As discussed in Section~\ref{uniform BKT}, BKT transition occurs at a critical phase-space density $D_c = \ln(380/\tilde{g})$, which for the experimentally relevant values of $\tilde{g}$ corresponds to $D_c \sim 6-10$. At that point density fluctuations are significantly suppressed, but cannot be completely neglected. This is one of the key reasons that makes the proper microscopic theory of the BKT transition in the Bose gas difficult.

In summary, $H_\theta$ can be used as an effective quantum Hamiltonian for the derivation of  some key features of long-range physics at low temperature, where $n_s\sim n$. However $H_\theta$ alone is not sufficient for the proper derivation of the BKT transition.

\subsection{Bogoliubov analysis}
\label{subsubsection:bogoliubov}

By performing Bogoliubov analysis near $T=0$ we can see more explicitly why the density fluctuations are suppressed at sufficiently low $T$, but the phase fluctuations are not, and also why it is natural to expect the low temperature state to be superfluid.

The basic idea is that we start with the assumption that the $T=0$ state of a weakly interacting gas is described by the uniform order parameter $\psi = \sqrt{n}\, e^{i\theta}$, find the excitation spectrum, and then consider the effects of the thermal occupation of the excitation modes. This approach may not seem justified in 2d, because in the end we find that thermal fluctuations destroy the order parameter, and thus invalidate our starting assumption. However we can qualitatively argue that it still works well as long as we have a well defined local order parameter, which is true if the order parameter is destroyed only at very large distances by the long-wavelength phase fluctuations. The applicability of the Bogoliubov approach to 2d quasi-condensates was formally justified in~\cite{Mora:2003,Castin:2004}.

Assuming contact interactions, the classical field Hamiltonian is given by
\begin{equation}
H = \frac{\hbar^2}{2m} \int  \left(\nabla \psi^*(\bs r)\right) (\nabla \psi(\bs r))\;d^2 r \;\
+ \ \frac{g}{2}\int  (\psi^*(\bs r))^2 (\psi(\bs r))^2  \;d^2 r \;  .
 \label{eq:classical_hamiltonian}
\end{equation}
The dynamics of $\psi(\bs r,t)$ are governed by the Gross-Pitaevskii equation~\cite{Gross:1961, Pitaevskii:1961}:
\begin{equation}
\left( -\frac{\hbar^2}{2m} \nabla^2 + g \left| \psi \right|^2 \right) \psi = i \hbar \frac{\partial \psi}{\partial t}  \; ,
\end{equation}
which we can derive from eq.~(\ref{eq:classical_hamiltonian}) by treating $\psi$ and $\psi^*$ as canonical variables.

We now define the local phase $\theta(\bs r,t)$ from $\psi(\bs r,t) = |\psi(\bs r,t)|\,e^{i\theta (\bs r,t)}$. Assuming that the density fluctuations are small, we write $|\psi(\bs r,t)|^2 =  n\, (1\, +\, 2\eta(\bs r,t))$, with $\eta \ll 1$ and $\int \eta \, d^2 r=0$. Within this approximation we get, up to the additive constant $gn^2L^2/2$,
\begin{equation}
H =  \frac{\hbar^2}{2m} n \int  (\nabla \theta (\bs r))^2 \;d^2 r \ +\
\int  \left[ \frac{\hbar^2}{2m}\,n\, (\nabla \eta(\bs r))^2 + 2gn^2 (\eta(\bs r))^2 \right]\;d^2  r \ ,
\end{equation}
and a set of coupled linear equations for the time evolution of $\theta(\bs r,t)$ and $\eta(\bs r,t)$:
\begin{eqnarray}
\frac{\partial \theta}{\partial t}& =&\ \frac{\hbar }{2m}\; \nabla^2 \eta\ -\ \frac{gn}{\hbar} (1 + 2\eta) \; , \label{eq:evol_theta}\\
\frac{\partial \eta}{\partial t} &=& - \frac{\hbar }{2m} \; \nabla^2 \theta \; .
\label{eq:evol_eta}
\end{eqnarray}

It is now convenient to Fourier expand the phase and the density fields:
\begin{equation}
\theta(\bs r,t) = \sum_{\bs k} c_{\bs k}(t)\, e^{i \bs k \cdot \bs r} \ , \quad
\eta(\bs r,t) = \sum_{\bs k} d_{\bs k}(t)\, e^{i \bs k \cdot \bs r} \ ,
\end{equation}
where $\bs k=2\pi (j_x,j_y)/L$, with $j_x,j_y$ integers, is a discrete variable since we consider for the moment a sample of finite size $L^2$. We will let $L\to \infty$ at the end of the calculation. The functions $\theta$ and $\eta$ are real, which implies $c_{\bs k}^*=c_{-\bs k}$ and $d_{\bs k}^*=d_{-\bs k}$. In addition the conservation of particle number $\int \eta\,d^2r=0$ leads to $d_0=0$. This yields the Hamiltonian
\begin{equation}
\label{bogoliubov in k}
H =   nL^2 \sum_{\bs k} \left[ \frac{\hbar^2 k^2}{2m} |c_{\bs k}|^2
+ \left( \frac{\hbar^2 k^2}{2m} + 2gn \right) |d_{\bs k}|^2 \right] \; ,
\end{equation}
and the coupled equations of motion (for $\bs k \neq 0$):
\begin{eqnarray}
\dot c_{\bs k}
&=& - \left( \frac{\hbar k^2}{2m} + \frac{2gn}{\hbar} \right) d_{\bs k}  \; ,\\
\dot d_{\bs k}
&=&   \frac{\hbar k^2}{2m} c_{\bs k} \; .
\end{eqnarray}
For $\bs k=0$ the equation of motion deduced from eq.~(\ref{eq:evol_theta}), $\dot c_0=-gn/\hbar$, simply gives the time evolution of the global phase of the gas.

In this formulation $c_{\bs k}$ and $d_{\bs k}$ are conjugate dimensionless quadratic degrees of freedom corresponding to the phase and the density fluctuations, respectively. At each $\bs k$ we have a harmonic oscillator-like Hamiltonian, and from the equations of motion we can read off the eigenfrequencies
\begin{equation}
\label{bogoliubov dispersion}
\omega_k = \sqrt{\frac{\hbar k^2}{2m} \left( \frac{\hbar k^2}{2m} + \frac{2gn}{\hbar} \right)} \; ,
\end{equation}
which is the well known Bogoliubov result. At low $k$, the eigenmodes are phonons with $\omega_k = ck$, where $c = \sqrt{gn/m}$. At high $k$ we have free particle modes with $\omega_k = \hbar k^2 /(2m) + gn/\hbar$. The crossover between the two regimes is at $k \sim 1/\xi = \sqrt{\tilde{g} n}$.

From this analysis, and specifically the results in eqs.~(\ref{bogoliubov in k}) and (\ref{bogoliubov dispersion}), we can draw several conclusions:

(1) From the dispersion relation $\omega_k$ we see that the excitation modes in this system have a non-zero minimal speed $c$. From the Landau criterion, we thus expect it to be superfluid with a critical velocity $v_c = c$. This argument is identical to the 3d case, and relies just on the existence of a (reasonably) well defined local order parameter, rather than true LRO. We note however that the Landau criterion is a necessary, but not a sufficient condition for  the existence of a superfluid state, because it does not address the question of metastability of the superfluid flow (see e.g. \cite{MaChap30:1985}).

(2) Both in the classical and in the quantum regime the harmonic oscillator at thermal equilibrium obeys the virial theorem, $\langle m \omega^2 x^2 \rangle = \langle p^2/m \rangle$, where $x$ and $p$ are position and momentum, respectively. Reading-off the coefficients in front of $|c_{\bs k}|^2$ and $|d_{\bs k}|^2$ in eq.~(\ref{bogoliubov in k}), we see that in our case this corresponds to:
\begin{equation}
\frac{\left\langle |d_{\bs k}|^2 \right\rangle}{\left\langle |c_{\bs k}|^2 \right\rangle} = \frac{\hbar^2 k^2/2m}{\hbar^2 k^2/2m\ +\ 2gn} \; .
\end{equation}
We thus explicitly see that long-wavelength phonons involve only phase fluctuations, since $\left\langle |d_{\bs k}|^2 \right\rangle \ll \left\langle |c_{\bs k}|^2 \right\rangle$ for $k \rightarrow 0$. On the other hand, the high-$k$ free particles involve both phase and density fluctuations in equal parts, since $\left\langle |d_{\bs k}|^2 \right\rangle = \left\langle |c_{\bs k}|^2 \right\rangle$ for $k \rightarrow \infty$.

(3) We also explicitly see that density fluctuations are not ``soft" Goldstone modes, because their energy cost does not vanish in the $k \rightarrow 0$ limit. The effect of interactions is to suppress (compared to the ideal gas) the density fluctuations at length scales $> \xi$. The fluctuations on short length scales ($< \xi$, corresponding to $k > 1/\xi$) are not suppressed by interactions, but those in any case do not have a diverging contribution.

(4) On the other hand, the energy cost of phase fluctuations (phonons) vanishes for $k \rightarrow 0$. This conclusion would be the same in 3d, but the crucial difference is that in 2d the density of states at low $k$ leads to a diverging effect of these fluctuations and the destruction of true LRO, as shown in the next subsection \S~\ref{sec:algebraic decay}.

(5) We can use this Bogoliubov analysis to provide an estimate of the relative density fluctuations $ \Delta n^2/n^2=4\langle \eta^2 \rangle$. In principle the thermal equilibrium distribution of the Bogoliubov modes is given by the Bose--Einstein distribution. For simplicity we approximate it in the following way; we suppose that the modes with frequency $\omega_k$ lower than $\kB T/\hbar$ have an average energy $\kB T/2$ (classical equipartition theorem) and that the higher energy modes have a negligible population. Therefore we take~\footnote{We use here that $c_{\bs k}$ and $d_{\bs k}$ are complex amplitudes, and that the modes $\bs k$ and $-\bs k$ are correlated, so that $d_{-\bs k}=d_{\bs k}^*$.}
 \begin{equation}
 nL^2 \left( \frac{\hbar^2 k^2}{2m} + 2gn \right) \langle |d_{\bs k}|^2\rangle = \left\{
 \begin{array}{l}
 \kB T/2\ , \quad \mbox{if}\ \hbar \omega_k < \kB T\ ,\\
  0\ , \qquad \quad\  \mbox{if}\ \hbar \omega_k > \kB T \ .
 \end{array}
 \right.
\label{}
\end{equation}
In addition we assume that $\kB T$ is notably higher than the interaction energy $gn$, which is the case in most experiments realized with cold atoms so far. The cutoff $\hbar \omega_k=\kB T$ then lies in the free-particle part of the Bogoliubov spectrum, at the wavevector $k_T\approx \sqrt{m\kB T}/\hbar$. We can now estimate the relative density fluctuations
\begin{equation}
\frac{\Delta n^2}{n^2}=4\sum_{\bs k} \langle |d_{\bs k}|^2\rangle
 \approx
\frac{L^2}{4\pi^2}\int_{k<k_T} \frac{4}{nL^2}
\frac{\kB T/2}{\hbar^2 k^2/2m\  +\ 2gn}\;d^2k
\label{}
\end{equation}
and we find
\begin{equation}
\frac{\Delta n^2}{n^2}\approx \frac{2}{n\lambda^2}\ln\left(\frac{\kB T}{2gn} \right)\;.
\label{eq:density_fluct}
\end{equation}
For realistic values of the ratio $\kB T/gn$ (i.e. not exponentially large), we recover here the previously announced result that density fluctuations are notably suppressed when $n\lambda^2 \gg 1$.

Based on these arguments, with some more quantitative justification we arrive at the same conclusions as before. First, for effectively describing the low  temperature state of the gas, the most important part of the Hamiltonian in eq.~(\ref{bogoliubov in k}) is the term corresponding to the phase fluctuations. Second, if we keep only that term, we also have to keep in mind that we are neglecting short-distance physics and effectively introducing a momentum cutoff at $k_{\rm max} = 1/\xi$. As before we can heuristically incorporate short-distance physics by replacing $n \rightarrow n_s$ to obtain:
\begin{equation}
\label{theta hamiltonian fourier}
H \approx n_s L^2 \sum_{\stackrel{\displaystyle{\scriptsize \bs k}}{ k<\xi^{-1}}} \frac{\hbar^2 k^2}{2m} |c_{\bs k}|^2=
\frac{\hbar^2}{2m}n_s \int (\bs \nabla \theta(\bs r))^2 \;d^2 r
 \; ,
\end{equation}
which coincides with the Hamiltonian $H_\theta$ given in eq.~(\ref{theta hamiltonian}).

\subsection{Algebraic decay of correlations}
\label{sec:algebraic decay}

To derive the low-$T$ behavior of the one-body correlation function $g_1(r)=\langle \psi^*(\bs r)\psi(0)\rangle$ at large distances, $r \gg \xi, \lambda$, we start with the Hamiltonian $H_\theta$, and the wave function with no density fluctuations $\psi(\bs r) = \sqrt{n_s}e^{i\theta(\bs r)}$. Note that this normalization of $\psi$ leads in $r=0$ to the incorrect value $g_1(0)=n_s$, whereas it should be $g_1(0) = n > n_s$. However, as discussed earlier, density fluctuations at short distances lead to a more complicated  decay of $g_1$, and at large $r$ replacing $n \rightarrow n_s$ is the appropriate normalization.

The long distance ($r\gg \xi$) behavior of $g_1(r)$ essentially depends on the population of phonon modes with wave vector $k\ll \xi^{-1}$, which coincides with the momentum cutoff introduced in $H_\theta$. The occupation of eigenmodes is in general given by the standard Bose result $(\exp(\beta \hbar \omega_k) - 1 )^{-1}$. Assuming as in eq.~(\ref{eq:density_fluct}) that $k_{\rm B}T>gn_s$, the phonon modes are in the regime $\hbar \omega_k \ll \kB T$ and the occupation number simplifies into $k_{\rm B} T / (\hbar \omega_k)$ which leads to (classical equipartition theorem)
\begin{equation}
\mbox{phonon modes:}\qquad n_sL^2 \frac{\hbar^2 k^2}{2m}\langle |c_k|^2\rangle=\frac{\kB T}{2}\;.
\label{}
\end{equation}
Introducing the real and imaginary part of the Fourier coefficients  $ c_{\bs k} = c_{\bs k}' + i c_{\bs k}''$, we have:
\begin{equation}
\label{c_k equipartition}
\left\langle |c_{\bs k}'|^2 \right\rangle = \left\langle |c_{\bs k}''|^2 \right\rangle = \frac{\pi}{n_s \lambda^2} \frac{1}{L^2 k^2} \; .
\end{equation}
We recall that $c_{\bs k}'$ and $c_{\bs k}''$ are independently fluctuating variables ($\left\langle c_{\bs k}' c_{\bs k}''\right\rangle = 0$) and that the mode $\bs k$ and $-\bs k$ are correlated because $\theta$ is real: $c'_{\bs k}=c'_{-\bs k}$ and  $c''_{\bs k}=-c''_{-\bs k}$.

We want to calculate:
\begin{equation}
g_1(r) = \left\langle \psi^*(\bs r) \psi(0) \right\rangle = n_s \left\langle e^{i(\theta(\bs r) - \theta(0))} \right\rangle \; ,
\end{equation}
where:
\begin{equation}
\theta(\bs r) - \theta(0) = \sum_{\bs k} c_{\bs k}' \left(\cos(\bs k \cdot \bs r) - 1 \right) - c_{\bs k}'' \sin(\bs k \cdot \bs r) \ .
\end{equation}
For each independent Gaussian variable $u$, $\left\langle e^{iu} \right\rangle = e^{- \frac{1}{2} \left\langle u^2 \right\rangle}$. Using eq.~(\ref{c_k equipartition}), and transforming the discrete sum over $\bs k$ into $L^2/(4\pi^2)\,\int d^2k$ we obtain:
\begin{equation}
\label{algebraic decay integral}
g_1(r)=
n_s \exp \left( - \frac{1}{2\pi n_s \lambda^2 } \int  \frac{1- \cos(\bs k \cdot \bs r)}{k^2}\;d^2 k \right) \; .
\end{equation}
The integral in the exponent has significant contributions only from modes $k > 1/r$ so that $1- \cos(\bs k \cdot \bs r) \sim 1$. Since we restrict our analysis to $r \gg \lambda$, this is not inconsistent with the classical field approximation which requires $k < 1/\lambda$. The upper limit of the integral is set by the short-distance cutoff $k_{\rm max} = 1/\xi$. We thus expect the integral to be $\sim \ln (r/\xi)$. More formally, we can note that $\nabla^2 \int (1- \cos(\bs k \cdot \bs r))k^{-2}\;d^2k = (2 \pi)^2 \delta(\bs r)$, from which we infer:
\begin{equation}
\int  \frac{1- \cos(\bs k \cdot \bs r)}{k^2}\;d^2k = 2\pi \ln \left( \frac{r}{\xi} \right) \; .
\end{equation}
This leads to:
\begin{equation}
\label{alebraic decay}
g_1(r) = n_s \left( \frac{\xi}{r} \right)^{1/(n_s \lambda^2)} \; .
\end{equation}
Note that depending on the relative size of $\lambda$ and $\xi$ we can also set the upper limit of the integral to $1/\lambda$, but this difference in the short-distance cutoff does not affect the main conclusion about the power law decay of correlations at large distances.

To summarize, we have shown that in an interacting 2d Bose gas at low $T$, the first order correlation function $g_1(r)$ decays algebraically with $r$ at large distances. The conclusion that $g_1(r)$ vanishes for $r \rightarrow \infty$ is consistent with the Mermin--Wagner theorem, i.e. the absence of BEC and true LRO at any non-zero $T$. However the decay of $g_1(r)$ is very slow and the system exhibits a ``quasi-long-range order". Further, as discussed in the following Section~\ref{uniform BKT}, the exponent $1/(n_s\lambda^2)$ is never larger than $1/4$ in the superfluid state, making the decay of $g_1(r)$ extremely slow. The superfluid state with suppressed density fluctuations can be viewed as a superfluid ``quasi-condensate", i.e. a condensate with a fluctuating phase \cite{Kagan:1987b,Popov:1987,Petrov:2004a}.


\section{The Berezinskii--Kosterlitz--Thouless (BKT) transition in a 2d Bose gas}
\label{uniform BKT}

Our analysis so far does not explain how the phase transition from the low temperature superfluid state to the high temperature normal state takes place. This transition is unusual because it does not involve any spontaneous symmetry breaking in the superfluid state, and in the usual classification of classical phase transition is termed ``infinite order", suggesting that most thermodynamic quantities (except for example superfluid density) vary smoothly at the transition. There is no true LRO on either side of the transition, but the functional form of the decay of $g_1(r)$ changes from algebraic in the superfluid state (corresponding to quasi-LRO) to exponential in the normal state (corresponding to no LRO).

The microscopic theory of the 2d superfluid transition was developed by Berezinskii~\cite{Berezinskii:1971} and Kosterlitz and Thouless~\cite{Kosterlitz:1973} (see \cite{Minnhagen:1987} for a more recent review).
The transition takes place in the degenerate regime, where the density fluctuations in an interacting gas are significantly suppressed. We therefore expect that the transition can still be at least qualitatively explained by considering only phase fluctuations. However, a sudden transition with a well defined critical point cannot be explained by considering only phonons, since we have seen in eq.~(\ref{alebraic decay}) that, while they destroy true LRO at any non-zero $T$, their effect grows smoothly with temperature.


\subsection{The role of vortices and topological order}

The key conceptual ingredient of the BKT theory is that in addition to phonons described by the Hamiltonian (\ref{theta hamiltonian}), another natural source of phase fluctuations are vortices, points at which the superfluid density vanishes, and around which the phase $\theta$ varies by a multiple of $2 \pi$. For our purposes we can consider only ``singly-charged" vortices with phase winding $\pm 2\pi$, which are energetically stable. Around an isolated single vortex, centered at the origin, the velocity field $\hbar \nabla \theta/m$ varies as $\hbar/(mr)$, corresponding to angular momentum $\hbar$ per particle. The two signs of the vortex charge correspond to the two senses of rotation around the vortex. The size of the vortex core (hole in the superfluid density) is set by the healing length $\xi$, so their presence is not inconsistent with the picture that the density fluctuations are suppressed at length scales $r \gg \xi$. In fact, it makes sense to speak of well defined individual vortices only if away from the vortex cores the density fluctuations are suppressed on the length scale $\xi$; otherwise we simply have a fully fluctuating thermal gas.

\begin{figure}[tbp]
\begin{center}
\includegraphics{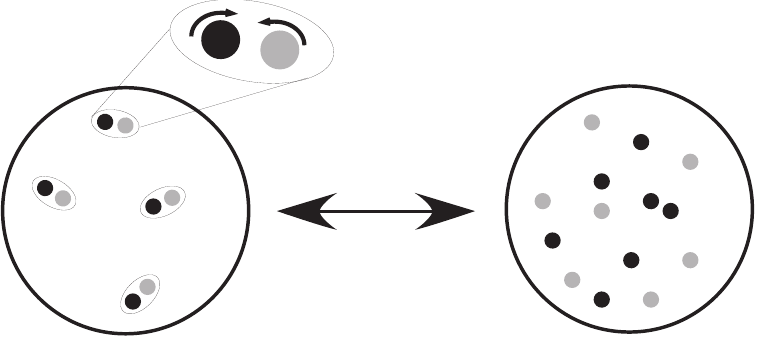}
\caption{The BKT mechanism at the origin of the superfluid transition. Below the transition temperature (left figure), vortices exist only in the form of bound pairs formed with vortices of opposite circulation. When approaching the transition point the density of pairs grows and the average size of a pair diverges. Just above the transition point (right figure), a plasma of free vortices is formed and the superfluid density vanishes.}
\label{fig:BKT mechanism}
\end{center}
\end{figure}

As we will illustrate below, once one considers vortices as another source of phase fluctuations, one can explain the microscopic mechanism behind the superfluid-to-normal phase transition. Below a well defined critical temperature $T_{\rm BKT}$, vortices can exist only in the form of bound (``dipole") pairs of vortices with opposite circulations $\pm 2 \pi$. Since they do not have any net charge, such pairs do not create any net circulation along closed contours larger than the pair size, which for a tightly bound pair is also of order $\xi$~\footnote{For a dipole field $|\bs \nabla \theta| \sim 1/r^2$, so the circulation $\oint \bs \nabla \theta\cdot d{\bs r}$ vanishes for large contours.}. Such pairs therefore have only a short-range effect on the phase $\theta$ and the velocity field, and do not have a large effect on the behavior of $g_1(r)$ at large distances. They fall under ``short-range physics" of the system and together with the residual density fluctuations they lead to some renormalization of $n_s$, but do not qualitatively alter the phenomenology of the long-range physics discussed in Section~\ref{sec: uniform 2d}.
On the other hand, above $T_{\rm BKT}$ unbinding of vortex pairs and proliferation of free vortices becomes energetically favorable. Free vortices then form a disordered gas of phase defects and completely ``scramble" the phase $\theta$ (see figure~\ref{fig:BKT mechanism}). This destroys the quasi-LRO and suppresses superfluidity. At even higher temperature where density fluctuations are strong, the notion of individual vortices becomes physically irrelevant.

In hindsight, we can associate superfluidity with presence of a ``topological order". Long-wavelength phase fluctuations (phonons) destroy true LRO, but do not alter the topology of the system. In other words, phonons lead to smooth local variations of the field $\psi$ which can be eliminated (or ``ironed out") by continuous deformations.
The same argument holds for bound vortex pairs which can be annihilated. Therefore, the superfluid quasi-condensate with no free vortices is topologically identical to the BEC with true LRO. On the other hand, an isolated free vortex cannot be unwound and eliminated from the system by continuous deformations of $\psi$; it affects the phase $\theta$ non-locally, at arbitrary large distances. The annihilation argument also does not work for a plasma of free vortices, because if we consider a closed contour of arbitrary large size it will in general not contain equal number of vortices with opposite charges. Therefore at any length scale the free vortex plasma is topologically different from an ordered BEC. Although we could not have necessarily anticipated this, we can deduce that topological order, rather than true LRO, is a sufficient condition for superfluidity in an interacting 2d Bose gas.


\subsection{A simple physical picture}

The full thermodynamic description of the 2d gas, including the role of vortices, is a difficuly task. It first requires to introduce the velocity fields~\cite{Nelson:1977} or the mass-current densities~\cite{Minnhagen:1981} in the fluid. The normal and superfluid components can then be extracted from the spatial correlation functions of this mass-current density~\cite{Minnhagen:1981}. Finally an analysis using renormalization group arguments leads to the so-called \emph{universal jump} for the superfluid density: This density takes the value $n_s=4/\lambda^2$ on the low temperature side of the transition point, and $n_s=0$ on the high temperature side.
The existence and the value of the universal jump in the superfluid density was formally derived by Nelson and Kosterlitz~\cite{Nelson:1977}, and it was first confirmed to high accuracy in experiments with liquid He films by Bishop and Reppy~\cite{Bishop:1978}.

This full description is outside the scope of this set of lectures. Here we  simply illustrate how vortices drive the BKT transition, starting from a superfluid with finite $n_s$, and then considering the free energy associated with spontaneous creation of a single free vortex.
Without any loss of generality, in order to simplify the calculations we consider a circular geometry, with $R \rightarrow \infty$ the radius of the system.
The kinetic energy cost of a single vortex placed at the origin is simply given by:
\begin{equation}
\label{vortex energy}
E = \int_\xi^R  \frac{1}{2} n_s \left( \frac{\hbar}{mr} \right)^2\;d^2 r  = \frac{\hbar^2 \pi}{m}n_s \ln \left( \frac{R}{\xi} \right) \; ,
\end{equation}
where as always we assume that only the superfluid component rotates under the influence of the vortex. The normal component does not have any phase stiffness and its motion is not affected by the presence of the vortex.

The entropy associated with a single vortex core is given by the number of distinct positions where a vortex of radius $\xi$ can be placed in a disc of radius $R$:
\begin{equation}
\label{vortex entropy}
S = k_{\rm B} \ln \left( \frac{R^2 \pi}{\xi^2 \pi} \right) = 2 k_{\rm B} \ln \left( \frac{R}{\xi} \right) .
\end{equation}
Note that in the above calculations we ignore the ``edge effects" such as the correction to the energy for an off-centered vortex. One can check that these effects are negligible for $R \gg \xi$.
Combining eqs.~(\ref{vortex energy}) and~(\ref{vortex entropy}), we get for the free energy $F= E - TS$:
\begin{equation}
\label{vortex free energy}
\beta F = \frac{1}{2} \left( n_s \lambda^2 - 4 \right) \ln \left( \frac{R}{\xi} \right) \; .
\end{equation}
We thus see that the free energy associated with a free vortex changes sign at $n_s \lambda^2 = 4$. Since $\ln(R/\xi)$ diverges with the size of the system, this point separates two qualitatively different regimes.
For $n_s \lambda^2 > 4$, $F$ is very large and positive, so  the superfluid is stable against spontaneous creation of a free vortex. On the other hand, for $n_s \lambda^2 < 4$, the large and negative $F$ signals the instability against proliferation of free vortices. Appearance of first free vortices reduces $n_s$  and makes the appearance of further free vortices even easier, and this avalanche effect renormalizes the superfluid density to zero. We thus find that in contrast to 3d, where below the BEC critical temperature the superfluid density grows smoothly, in 2d the superfluid density cannot have any value between $4/\lambda^2$ and 0.

Even though it does not explicitly address the microscopic origin of the vortex (the breaking of a vortex pair), this simple calculation correctly predicts the result for the universal jump in the superfluid density which takes place at the transition temperature $T_{\rm BKT}$:
\begin{equation}
\label{universal jump}
n_s \lambda^2 = 4 \; .
\end{equation}
This success relies on the fact that the above derivation is a powerful self-consistency argument. Whatever the origin of the vortex, and the relation between $n_s$ and total density $n$, it shows that it is inconsistent to suppose that we have a system with superfluid density which is non-zero, but smaller than $4/\lambda^2$. Remarkably, this result also does not depend on the strength of interactions $\tilde{g}$, even though we know that the phase transition is mediated by interactions, since it does not occur in the ideal gas. Recent classical field Monte-Carlo calculations performed with parameters relevant for atomic gases~\cite{Simula:2006,Giorgetti:2007,Bisset:2009b} have confirmed the proliferation of vortices around the critical point characterized by eq.~(\ref{universal jump}), although the transition was rounded off by finite-size effects.

If we repeat the above arguments for tightly bound vortex pairs, we find that a finite density of pairs is present in the gas at any non-zero temperature. The energy of a pair is finite since the velocity field decays as $v \propto 1/r^2$ and the integral $\int  v^2\;d^2 r $ is convergent. On the other hand, the entropy is still divergent and essentially identical to the result of eq.~(\ref{vortex entropy}), since the size of a tightly bound pair is of the same order as the size of a single vortex. The free energy for vortex pairs is therefore always negative. At any non-zero $T$ pairs are continuously created and annihilated through thermal fluctuations.

As the temperature is increased, but still kept below  $T_{\rm BKT}$, the density of pairs grows and also thermal fluctuations result in pairs of increasing size. As the average size of the pairs becomes comparable to the distance between the pairs, they start to overlap and this leads to effective screening of the attraction between two bound vortices, making it easier for fluctuating pairs to grow to even larger sizes. We can use an analogy with a Coulomb gas: Two nominally paired but well separated vortices create a field which polarizes the more tightly bound vortex dipoles between them. This results in an effective dielectric constant which reduces the attraction between two oppositely charged vortices. As $T_{\rm BKT}$ is approached from below, this creates an avalanche effect which eventually leads to breaking up of pairs and creation of a plasma of free vortices. Within the Coulomb gas analogy, this plasma provides a perfect screening at the transition point: A charge added at a given point does not change the flux of the electric field across a large radius circle centred on this point. Proper BKT calculation identifies the phase transition with the temperature at which the average size of the pairs, or equivalently the screening dielectric constant, diverges.


\subsection{Results of the microscopic theory}
\label{subsec:BKT microscopic results}

The relation (\ref{universal jump}) between the superfluid density and the  temperature $T_{\rm BKT}$ at the critical point is elegant and universal, in the sense that it does not depend on the interaction strength $\tilde{g}$. However this self-consistent result alone does not allow us to predict the value of $T_{\rm BKT}$ in a given system. It just tells us that whatever $T_{\rm BKT}$ is,  the superfluid density jumps accordingly to $4/\lambda^2$ at the transition.

Calculating the actual value of $T_{\rm BKT}$ in terms of the bare system properties $n$ and $\tilde{g}$ is a difficult problem, because it depends on the short-distance physics, such as density fluctuations which control the relationship between $n_s$ and $n$ at the transition. In the weak coupling limit, $\tilde{g} \ll 1$, a combination of analytical~\cite{Fisher:1988} and numerical \cite{Prokofev:2001,Prokofev:2002} efforts gives the value for the critical phase-space density:
\begin{equation}
\label{D_c PS}
D_c = (n \lambda^2)_c = \ln \left( \,C\,/\,\tilde{g} \,\right) \; ,
\end{equation}
where the dimensionless constant $C = 380 \pm 3$ is obtained by a classical field Monte-Carlo simulation~\cite{Prokofev:2001}. The calculation leading to eq.~(\ref{D_c PS}) is formally valid only in the weak coupling limit $\tilde{g} \ll 1$.
We can set one obvious bound on its validity by noting that at the transition $n \geq n_s$. Setting $D_c > 4$ we obtain $\tilde{g} \leq  7$. This result is remarkably close to our estimate of the strong coupling limit $\tilde{g} = 2 \pi$ (eq.~(\ref{strong g})).

The numerical simulation of~\cite{Prokofev:2001} provides another quantity of interest, which characterizes the reduction of density fluctuations due to interactions. The authors of~\cite{Prokofev:2001} introduce the \emph{quasi-condensate density} defined as:
\begin{equation}
n_{\rm qc} \equiv \left(2 n^2 - \left\langle n^2(\bs r) \right\rangle \right)^{1/2}\ .
\end{equation}
When interactions are negligible, $\langle n^2\rangle=2\langle n\rangle^2$ and
$n_{\rm qc}=0$. On the other hand if density fluctuations are completely suppressed, $\langle n^2\rangle=\langle n\rangle^2$ and $n_{\rm qc}=n$. According to~\cite{Prokofev:2001}, at the critical point
\begin{equation}
 \frac{n_{\rm qc}}{n}=\frac{7.16}{\ln (C/ \tilde g)}\ .
\label{}
\end{equation}
 This result indicates that $n_{\rm qc}$ is of the order of the total density $n$ at the transition point, unless the interaction strength $\tilde g$ is exponentially small. In other words, for realistic parameters density fluctuations are notably reduced in the vicinity of the BKT transition, which justifies the simplified Hamiltonian (\ref{theta hamiltonian}) used above. Actually this result sets a stronger constraint than eq.~(\ref{D_c PS}) on the applicability of the classical Monte-Carlo analysis: the condition $n \geq n_{\rm qc}$ requires $\tilde g \leq 0.3$.

Note that the terminology \emph{quasi-condensate density} can sometimes be misleading. The quantity $n_{\rm qc}$ takes a non-zero value even above the critical temperature for the BKT transition. It refers only to the properties of the density distribution in the gas, and not to the phase distribution as the word \emph{condensate} might suggest. For example, we can have $n_{\rm qc}\sim n$ at $T>T_{\rm BKT}$, but this does not imply that a large contrast interference would be observed if one would superpose a pair of 2d gases prepared in the non-superfluid regime (see Section~\ref{sec:experimental probes}).

Finally, we note that as the transition temperature is approached from above, the length scale $\ell$ characterizing the exponential decay of correlations $g_1(r) \sim e^{-r/\ell}$ in the normal state diverges as:
\begin{equation}
\label{BKT diverging ell}
\ell = \lambda \exp\left(\frac{ \sqrt{a\,T_{\rm BKT}}}{\sqrt{T-T_{\rm BKT}}}
 \right)
\; ,
\end{equation}
where $a$ is a model-dependent dimensionless constant. Diverging correlation length is a very general property of phase transitions, but while in the case of most conventional (3d) second-order phase transitions the divergence of the correlation length is polynomial, in the case of the BKT transition it is exponential. This makes the critical region above $T_{\rm BKT}$ larger, and has implications for the broadening of the transition in finite size systems (see e.g. \S~\ref{width of the crossover in a box}).


\section{The 2d Bose gas in a finite box}
\label{sec:2d gas in a box}

It is well known that finite size effects can play a significant role in the quantitative analysis of the phase transitions that are observed experimentally. In the two-dimensional situation of interest here, this is even more the case because the thermodynamic limit is reached only when $\ln(R/\xi) \gg 1$ (see for example eq. (\ref{vortex free energy})), which can be only marginally true for realistic systems. The analysis of these finite-size effects is therefore crucial for understanding the observed phenomena, and we will do so for the cases of a flat box potential (this section) and a harmonic trap (next section).

We consider in this section a gas of $N$ particles confined in a flat box of area $L^2$.
The box is supposed to be square ($L_x=L_y=L$) except in the last subsection (\S~\ref{subsect:anisotropic_flat_potential}) where we discuss possible effects due to an anistropic confinement ($L_x \neq L_y$). The confinement introduces a natural energy scale $E_0=\hbar^2/(mL^2)$ in the problem and makes it possible to reach a true Bose--Einstein condensate at non-zero temperature, in contrast to the infinite case. In this section we first review the results that can be derived for the ideal gas case, and then discuss what happens for an interacting system. We will assume that the size $L$ is much larger than the thermal wavelength $\lambda$ so that  $E_0 \ll k_{\rm B} T$.

\subsection{The ideal Bose gas}
\label{subsec:ideal_Bose_gas_box}

In absence of interactions, the statistical description of a Bose gas in a square box of size $L$ is straightforward. Let us choose for simplicity periodic boundary conditions so that the single particle eigenstates are plane waves $e^{i\bs k \cdot \bs r}/L$, of energy $\epsilon_k=\hbar^2k^2/2m$, where the momentum $\bs k=(j_x,j_y) (2\pi/L)$, with $j_{x,y}$ positive, zero or negative integers. The chemical potential $\mu$ is always negative so that the fugacity $Z=e^{\beta\mu}$ lies in the interval $0<Z<1$. Three regimes can be identified, the first two being identical to what we have met for the infinite case:
\begin{itemize}
 \item
The non-degenerate, high temperature regime with a phase space density $D=n\lambda^2 \ll 1$. This corresponds to a negative chemical potential such that $|\mu|\gg k_{\rm B}T$ ($Z\ll 1$). The one-body correlation function is a gaussian function in this regime, $g_1(\bs r)=n\,e^{-2\pi r^2/\lambda^2}$, and is vanishingly small for distances $r \gg \lambda$, which are much smaller than the box size $L$. The confinement has no significant consequence on the coherence of the gas in this regime.

 \item
The degenerate, but non-condensed regime, where the momentum distribution is bimodal, with a Lorentzian shape for small $k$ ($k\lambda \ll \sqrt{4\pi}$) and a gaussian shape for large $k$. The one-body correlation function decays exponentially at large $r$, with a characteristic decay length $\ell=\lambda e^{D/2}/\sqrt{4\pi}$. No significant condensed fraction appears as long as $\ell$ is small compared to the size $L$ of the sample, i.e. when
$D<\ln(4\pi L^2/\lambda^2)$ or equivalently $|\mu|\gg E_0/2 $.

\item
The condensed regime, which occurs when the characteristic decay length $\ell$ of $g_1$ is larger than the system size $L$. This occurs when the phase space density $D$ reaches the value $\ln(4\pi L^2/\lambda^2)$ (or equivalently $|\mu| \leq E_0/2$). A significant phase coherence then exists between any two points in the gas.
\end{itemize}


\subsection{The interacting case}

We now turn to the interacting case with repulsive interactions and discuss what can be expected in the vicinity of the BKT transition. For now we assume that the size of the sample is large enough so that $D \ll \ln(4\pi L^2/\lambda^2)$ at the point where phase space density $D$ is equal to $D_c$, the critical phase space density for the BKT transition in an infinite system (see eq. (\ref{D_c PS})). This condition has the following physical meaning: suppose that we increase the density of particles $n$ at fixed temperature; the point where the superfluid transition occurs is reached \emph{well before} the point at which the Bose--Einstein condensation due to the finite system size would occur in absence of interactions (see section \ref{subsec:ideal_Bose_gas_box} above).

We first recall the nature of the superfluid transition in an infinite, homogenous sample. When $D$ is notably below $D_c$, but larger than 1, one expects that no superfluid component is present and $g_1(r)$ decays exponentially. When $D$ reaches $D_c$ the superfluid transition occurs and $g_1(r)$ decays algebraically (eq. (\ref{alebraic decay})): $g_1(r) \approx n_s(\xi/r)^{\alpha}$ for $r>\xi$, with $\alpha=1/(n_s\lambda^2)$ and $n_s\lambda^2 \geq 4$. According to the Penrose--Onsager criterion, no condensate is expected in an infinite system since $g_1(r)$ vanishes at infinity for any non-zero temperature. In sharp contrast with the infinite case, we now show that the BKT transition in a realistic finite system is always accompanied by the appearance of a significant condensed fraction, defined as the largest eigenvalue $\Pi_0$ of the one-body density matrix. The basic reason for this effect is that the algebraic decay of $g_1(r)$ is extremely slow. To prove this result we proceed in two steps: first we give a general relation between $\Pi_0$ and the value of $g_1(r)$ for distances $r$ comparable to the size $L$ of the box; then we discuss what a realistic value of $\Pi_0$ can be for a typical atomic gas.

Let us denote by $\Pi_j$ and $\phi_j(\bs r)$ the eigenvalues and eigenstates of the one-body density matrix. The condensed fraction $\Pi_0$ is associated with the eigenstate $\phi_0(\bs r)=1/L$. We now consider the general expansion of $g_1$
 \begin{equation}
g_1(\bs r)=N\sum_j \Pi_j \phi_j^*(0)\phi_j(\bs r)
\label{}
\end{equation}
and integrate this expression over the area $L^2$ centered on the origin. For simplicity we integrate the left hand side over a disk of radius $R=L/\sqrt \pi$ and the right hand side over a square of side $L$. This simplification cannot significantly affect our conclusions. The left hand side gives
 \begin{equation}
\int g_1(\bs r)\;d^2r= 2\pi \,n_s\, \int_{\xi}^{R} (\xi/r)^{\alpha} \,r\,dr \simeq L^2 \,\frac{2}{2-\alpha}\,\pi^{\alpha/2}\,g_1(L) \; .
\label{}
\end{equation}
On the right hand side, only the contribution of $j=0$ is non zero and gives $N\Pi_0$. All the $\phi_j$'s with $j\neq0$ are orthogonal to $\phi_0$, so their integral over $L^2$ is zero. We thus get
$g_1(L)/n \sim  \Pi_0$. Using eq. (\ref{healing length}) we can also write this result for the condensed fraction as $\Pi_0 \sim (n_s/n) \tilde g^{-\alpha/2} N^{-\alpha/2}$.

For $\alpha \leq 1/4$, in practice we have $\tilde g^{-\alpha/2} \sim 1$ and $n_s\sim n$, so just below the transition temperature $\Pi_0 \sim N^{-1/8}$. Taking $N=10^5$ as a typical value for cold atom experiments, we get $\Pi_0 \sim 0.25$. \footnote{ We get an equivalent estimate from $\Pi_0 \sim g_1(L)/n$ and $n_s \sim n$ in the algebraic decay regime. In cold atom gases, the typical values of $\lambda$ and $\xi$ are $0.1 - 1\;\mu$m, while the maximal system size is $L\sim 100\;\mu$m. At the transition point we have $\Pi_0 \sim (10^{-3})^{1/4}$ to $(10^{-2})^{1/4}$ $\sim$  $0.2$ to $0.3$.}. We therefore meet here a paradoxical situation: the appearance of a non-zero condensed fraction may be used as a signature of the BKT transition, whereas the BKT mechanism was presented (for an infinite system) as a feature that takes place in a 2d interacting gas \emph{instead of} the usual BEC of 3d Bose fluids. Note that in a ``true" BEC the condensed fraction $\Pi_0$ should not explicitly depend on $N$. However, for $\alpha \leq 1/4$ this distinction becomes experimentally irrelevant, and in order to observe a BKT transition with no significant BEC one would need to consider unrealistically large samples. This was pointed out by the authors of \cite{Bramwell:1994} who wrote the famous statement (in the context of 2d magnetism): ``With a magnetization at the BKT critical point smaller than 0.01 as a reasonable estimate for the thermodynamic limit, the sample would need to be bigger than the state of Texas for the Mermin--Wagner theorem to be relevant!". A similar remark holds in the context of superfluid helium films~\cite{Bishop:1978}.


\subsection{Width of the critical region and crossover}
\label{width of the crossover in a box}

The intricate mixing between the BKT mechanism and the emergence of a significant degree of coherence exists even at temperatures slightly above the BKT transition point. We mentioned in \S~\ref{subsec:BKT microscopic results} that in the normal state the characteristic decay length $\ell$ of $g_1(r)$ diverges exponentially in the vicinity of the critical point: $\ln(\ell/\lambda) \approx (aT_{\rm BKT}/(T-T_{\rm BKT}))^{1/2}$, where $a$ is a model-dependent coefficient (see eq. (\ref{BKT diverging ell})). The critical region where $\ell$ becomes larger than $\lambda$ as a precursor of the BKT transition is therefore very broad, $(T-T_{\rm BKT}) \sim T_{\rm BKT}$. Further, there clearly exists a temperature close to (but still above) $T_{\rm BKT}$ for which $\ell$ exceeds the system size. At this temperature a significant condensed fraction appears in the system. Because of the exponential variation of $\ell$ with $T-T_{\rm BKT}$, the temperature range where $\ell \gtrsim L$, and condensation gradually sets in, can be significant~\footnote{Here we can assume that $\lambda$ under the logarithm on the right hand side is constant over the range $\Delta T$.}:
\begin{equation}
\label{bec crossover width}
\frac{\Delta T}{T_{\rm BKT}}=\frac{\Delta D}{D_c}\sim \frac{a}{(\ln(L/\lambda))^2} \; .
\end{equation}

Although cold atom systems are so far commonly confined in harmonic rather than box-like potentials, it is interesting to provide an estimate for this case. Taking $a=1$ and  reasonable values for $L/\lambda$ between 10 and 100,
the BKT transition is expected to become a crossover with a relative width $\Delta T / T_{\rm BKT} $ ranging from 5\% to 20\%.

The above analysis explicitly concerns only the emergence of a non-zero condensed fraction, but it also suggests broadening of the universal jump in the superfluid density. For example, using standard Bogoliubov argument (see \S~\ref{subsubsection:bogoliubov}) we can deduce that finite condensed fraction in an interacting system also implies a finite superfluid density. Also, if we take the definition of superfluid density which associates it with the energy cost of twisting the phase of the wave function at the edge of the system \cite{Leggett:1973,Fisher:1973}, we again conclude that $\ell \gtrsim L$ implies a finite superfluid density. In the critical region, quantitative conclusions might actually depend on what theoretical definition of superfluid density we accept, but the qualitative conclusions will not change.


\subsection{What comes first: BEC or BKT?}
\label{what comes first in a box}

This is an often raised and subtle question. We have so far discussed the case of a large system such that $D_c \ll \ln(4\pi L^2/\lambda^2)$ where $D_c=\ln(380/\tilde g)$ is the critical phase space density for the BKT transition in an infinite system. For such large systems, the first relevant mechanism that occurs when increasing the phase space density is a BKT transition. However we have seen that the approach of the BKT threshold always results in the appearance of a significant condensed fraction. We can qualify this ``BKT-driven" condensation as ``interaction-enhanced", since it would not take place in an ideal gas with the same density and temperature.
Experimentally, the strength of interactions in cold atom systems can be dynamically controlled using a Feshbach resonance~\cite{Tiesinga:1993,Inouye:1998}. One can therefore imagine preparing a non-interacting gas at phase space density where no condensation occurs and then driving the condensation via the BKT mechanism by turning on the interactions. In the opposite regime of a small system where the BKT transition would require $D_c > \ln(4\pi L^2/\lambda^2)$, the first phenomenon that is encountered when the phase space density is increased is ``conventional" Bose--Einstein condensation as for an ideal Bose gas. As in the 3d case, in presence of weak repulsive interactions the formation of a condensate is accompanied by the apparition of a superfluid fraction with a comparable value.

It would be very interesting to study cold atomic gases in a (quasi-)uniform potential and vary the experimental parameters so as to explore both regimes.
However we point out that this will not be an easy task if we require that the criteria for the two types of transitions are well separated, for example by more than the crossover width discussed in \S~\ref{width of the crossover in a box}.
To analyze the system requirements for reaching the two different regimes, it is convenient to fix the ratio $L/\lambda$ and write $\ln(4\pi L^2/\lambda^2) = \gamma D_c$. Now $\gamma$ is a dimensionless parameter such that $\gamma > 1$ means that (as particle number is increased) condensation occurs via the BKT mechanism. The critical number for the BKT transition is then
\begin{equation}
N_{\rm c} = \frac{L^2}{\lambda^2} D_c = \frac{1}{4\pi} D_c e^{\gamma D_c} \; ,
\end{equation}
while the critical number for condensation in an ideal gas is $\gamma N_{\rm c}$. For illustration purposes we may define the BKT regime by $\gamma \geq 1.5$, and the BEC regime by $\gamma \leq 0.5$. (For values of $\gamma$ close to 1, the two effects are difficult to disentangle experimentally.) The experiments with cold atoms have so far been performed at coupling strength $\tilde{g} \sim 10^{-2} - 10^{-1}$. Taking $\tilde{g} = 0.1$, $\gamma=1$ corresponds to $N_{\rm c} \approx 2.5 \times 10^3$, and the BKT regime $\gamma =1.5$ corresponds to $N_{\rm c} \approx 1.5 \times 10^5$, which is easily achievable. However the opposite BEC regime of $\gamma = 0.5$ corresponds to $N_{\rm c} \approx 40$; studying such a small particle number is experimentally very challenging, although it might become feasible with the development of single-atom detection~\cite{Schrader:2004,Nelson:2007,Bakr:2009}. For a more weakly interacting gas with $\tilde{g} = 0.01$, $\gamma=1$ corresponds to $N_{\rm c} \approx 3 \times 10^4$, and $\gamma =0.5$ to $ N_{\rm c} \approx 160{}$, which might be easier to explore. On the other hand $\gamma =1.5$ corresponds to $N_{\rm c} \approx 6 \times 10^6$, which would be experimentally challenging. It would therefore generally be difficult to explore both the large (BKT) and the small (BEC) system regime using the same value of $\tilde{g}$, and reaching the BEC regime may require a more weakly interacting quasi-2d atomic gas than has so far been studied.


\subsection{The case of anisotropic samples}
\label{subsect:anisotropic_flat_potential}

So far we have assumed that for phase space densities $D$ larger than the critical value $D_c$ for the BKT transition, the functional form of $g_1$ found in the infinite case (algebraic decay) remained valid for a finite size system. This assumption is reasonable for square samples ($L_x=L_y$), but may not be valid for anisotropic samples, with a width along one direction (say $x$) much larger than the other one: $L_x\gg L_y$. We briefly review the expected properties in this regime, which is relevant for several of the previous or current experimental setups. Since we are interested in the regime $D>D_c$, we assume that a superfluid component is present in the sample and we use the hamiltonian $H_\theta$ given in (\ref{theta hamiltonian fourier}) to estimate the amplitude of phase fluctuations and their consequence on the one-body correlation function $g_1$.

We start from the result (\ref{algebraic decay integral}) obtained in the infinite case. In a finite size system the integral over $\bs k$ is replaced by a discrete sum over $\bs k=2\pi(j_x/L_x,j_y/L_y)$ times the constant prefactor $4\pi^2/(L_x L_y)$:
 \begin{equation}
\ln (g_1(\bs r)/n_s)=-\frac{2\pi}{n_s\lambda^2 L_xL_y}\sum_{\bs k}\frac{1-\cos(\bs k\cdot\bs r)}{k_x^2+k_y^2} \ .
\label{eq:discretized_g1}
\end{equation}
We are interested here in the decay of $g_1$ along the long axis of the sample and we choose $\bs r=x \bs u_x$, where $\bs u_x$ is the unit vector along the $x$ direction. We  take $x\ll L_x$ so that the finiteness of the sample along $x$ has no relevance here. On the contrary the size $L_y$ ($\ll L_x$) will play an important role since we can choose $x$ either small or large compared to $L_y$. We now show that these two cases lead to different decaying regimes for $g_1$.

Since we assume $|x|\ll L_x$, the discrete sum over $k_x$ can always be replaced by an
integral  and we get:
\begin{equation}
\ln (g_1(x)/n_s)=-\frac{1}{n_s\lambda^2 L_y}\sum_{k_y}\int \frac{1-\cos(k_x x)}{k_x^2+k_y^2}dk_x=-\frac{\pi}{n_s\lambda^2 L_y}\sum_{k_y}\frac{1-e^{-|xk_y|}}{|k_y|}
\ .
\label{}
\end{equation}
We single out the contribution of $k_y=0$ in the sum, introduce a cutoff $k_{\rm max}$ for large $k_y$ and use $\sum_{j=1}^{j_{\rm max}}1/j\approx \ln j_{\rm max}$
and $\sum_{j=1}^{\infty}\zeta^j/j=-\ln(1-\zeta)$ for $0<\zeta<1$. We then get
\begin{equation}
\ln (g_1(x)/n_s)=-\frac{1}{n_s\lambda^2}
\left\{
\frac{\pi |x|}{L_y} +\ln \left[
\frac{k_{\rm max}L_y}{2\pi}\left(1-e^{-2\pi\, |x|/L_y} \right)
\right]
\right\}\ .
\label{eq:g1_elongated}
\end{equation}
Two regimes clearly appear in this expression. If $\pi|x|\ll L_y$ then the logarithm on the right hand side is the dominant term, and we recover the algebraic decay that holds for an infinite system:
\begin{equation}
k_{\rm max}^{-1}\ll |x|\ll L_y:\quad g_1(x)\approx \frac{n_s}{(|x|k_{\rm max})^{\alpha}}
 \quad \mbox{with}\quad
 \alpha=\frac{1}{n_s\lambda^2}\ .
\label{}
\end{equation}
This result is intuitive: as long as we probe the coherence of the system on a distance shorter than the smallest size of the sample, the anisotropy introduces no significant deviation with respect to an infinite system.

The situation is dramatically different when $|x|\gg L_y$. In this case the dominant contribution on the right hand side of (\ref{eq:g1_elongated}) is the linear term $\pi |x|/L_y$. This term, which originates from the contribution of the $k_y=0$ mode to the sum (\ref{eq:discretized_g1}), leads to an exponential decay of $g_1$:
\begin{equation}
  L_y\ll |x|:\quad g_1(x)\approx n_s\,e^{-|x|/d}
 \quad \mbox{with}\quad
 d=n_s\lambda^2 L_y/\pi\ .
\label{}
\end{equation}
It is quite remarkable that when we probe the coherence of this anisotropic system on distances $x\geq d>L_y$, we obtain an exponential decay as if the system was not superfluid. At the same time, the characteristic distance over which $g_1$ decays, $d$, explicitly depends on the superfluid density. The physical interpretation of this counterintuitive result is that over such distances the system acquires a quasi-one dimensional character: the phase stiffness between the origin and the point at coordinate $x$ is decreased with respect to an infinite plane because there is a severe reduction in the number of independent paths connecting these two points. Ultimately, for very large $|x|$, only the channel $k_y=0$ contributes significantly to the connection between these two points. This explains why, although the system is superfluid, the decay of $g_1$ turns to an exponentially decaying function, that is characteristic of 1d degenerate gases (see also~\cite{Gerbier:2004b} for a similar discussion for elongated atomic 3d gases).


\section{The 2d Bose gas in a harmonic trap}
\label{sec:2d gas in a trap}

Up to now, experiments performed with (quasi-)2d atomic gases used harmonic trapping in the $xy$ plane. We discuss in this section how the presence of the harmonic trapping potential modifies the above results. We shall see that it can lead to a dramatic change of the properties of the system, through the modification of the density of states of the single particle Hamiltonian. In particular, in this case ``conventional" Bose--Einstein condensation, as defined through the saturation of excited states at some non-zero temperature $T_{\rm c}$, can occur in the ideal Bose gas even in the thermodynamic limit. However, in presence of repulsive interactions, and in large enough systems, this type of condensation is suppressed and replaced by the BKT normal to superfluid transition.


\subsection{The ideal case}

Consider for simplicity an isotropic 2d harmonic potential $V(r)=m\omega^2r^2/2$. The single particle energy levels are $E_j=(j+1)\hbar \omega$, with $j$ positive or zero integer, and each level having a degeneracy $g_j=j+1$. The maximum number of atoms $N_{\rm c}$ that can be placed in all excited states ($j>0$) at a given temperature $T$ is obtained by choosing a chemical potential $\mu$ equal to the ground state energy:
 \begin{equation}
 \NctwoDid(T) = \sum_{j=1}^{+\infty} \frac{g_j}{e^{\zeta j}-1}\ ,
\label{eq:Nc_harm}
\end{equation}
where $\zeta=\hbar \omega/(k_{\rm B} T)$. Assuming $\zeta \ll 1$, the discrete sum can be replaced by an integral and one obtains
 \begin{equation}
 \NctwoDid(T) \approx \frac{\pi^2}{6} \,\left(
 \frac{k_{\rm B}T}{\hbar \omega}
 \right)^2\ .
\label{eq:NC_harm2}
\end{equation}
For an atom number $N>N_{\rm c}(T)$ there must be at least $N-N_{\rm c}$ atoms occupying the single particle ground state $j=0$. Equivalently, for a given atom number $N$ placed in the trap, there must be a significant fraction of the atoms that occupy the ground state $j=0$ if the temperature is reduced below the critical value
 \begin{equation}
k_{\rm B} T_{\rm c}  =\frac{\sqrt 6}{\pi} \hbar \omega \sqrt N\ .
\label{eq:Tc_harm}
\end{equation}

Since the ground state is separated from the first excited state by a non-zero gap $\hbar \omega$, this Bose--Einstein condensation can be viewed as a natural consequence of the finite size of the system~\cite{Bagnato:1987}, similar to the condensation of the ideal gas in a finite box. However, one can see that the condensation of the ideal gas in a 2d harmonic trap is a more interesting phenomenon by considering the appropriately defined thermodynamic limit for the harmonic confinement. This limit is obtained by taking $N\to \infty$ and $\omega\to 0$, while keeping $T$ and $N\omega^2$ constant. For a gas described by Boltzmann statistics, this ensures that the central density $n_0=Nm\omega^2/(2\pi k_{\rm B} T)$ remains constant. Eq. (\ref{eq:Tc_harm}) leads to a non-zero critical temperature in the thermodynamic limit, contrarily to what happens in the uniform case. This result can be understood by noticing that the density of states has a different functional form in a box ($\rho(E)$ constant) and in a 2d harmonic potential ($\rho(E) \propto E$) \cite{Bagnato:1991}. The vanishing density of states at $E=0$ for a 2d harmonic potential leads to a similar situation to the 3d uniform case, hence the possibility for a genuine Bose--Einstein condensation in the ideal gas (for a discussion of small logarithmic anomalies in the compressibility of the trapped ideal 2d gas, see \cite{Yukalov:2005}).

Quite remarkably the result (\ref{eq:NC_harm2}) can be recovered by starting from the uniform result given in eq.~(\ref{no saturation}): $n=-\lambda^{-2}\ln (1-e^{\beta \mu})$ and using a local density approximation (LDA). This approximation amounts to replacing the uniform chemical potential $\mu$ by the local one $\mu-V(r)$, which gives the following expression for the total atom number:
 \begin{eqnarray}
N &=&-\lambda^{-2} \int \ln (1-e^{\beta (\mu-V(r))})\;2\pi r\;dr  \nonumber \\
 &=&-\left(\frac{k_{\rm B}T}{\hbar \omega} \right)^2\int_0^{+\infty}
\ln\left( 1-Ze^{-R^2/2}\right)\,R\;dR
\label{eq:ideal_LDA}
\end{eqnarray}
where we set $R=r/r_T$ with $r_T^2=k_{\rm B} T/m\omega^2$. For $\mu=0$, the result coincides with eq.~(\ref{eq:NC_harm2}). Therefore in spite of the fact that LDA leads to a diverging spatial density at the center of the trap for $\mu=0$ [$n(r)\propto -\ln (r)$], it provides the same upper bound $N_{\rm c}$ as eq.~(\ref{eq:Nc_harm}) for the total number of atoms, assuming no macroscopic occupation of the single particle ground state.


\subsection{LDA for an interacting gas}

In order to take into account repulsive interactions for a trapped gas, we use again the local density approximation. We will start with an analytical mean-field  treatment based on the Hartree--Fock approximation. We will then use the numerical results of a classical field Monte-Carlo approach \cite{Prokofev:2002} that will provide a more precise determination of the BKT transition.

In the mean-field Hartree--Fock approach when no condensate is present, interactions are taken into account by adding the energy $2gn(r)$ to the external potential \cite{Pitaevskii:2003,Pethick:2002}. The local chemical potential is now
$\mu-V(r)-2g n(r)$ so that the local phase space density $D(r)=n(r)\lambda^2$ is the solution of the implicit equation
 \begin{equation}
D(r)=-\ln\left\{ 1-Z\exp[-\beta V(r)-\tilde g D(r)/\pi] \right\}\;.
\label{eq:interact_2d_trap}
\end{equation}
Putting $R=r/r_T$ as above, we can write the total atom number as
\begin{equation}
\frac{N}{\NctwoDid}=\frac{6}{\pi^2} \int_0^{+\infty}
  D(R)\,R\,dR
\label{eq:interact_N}
\end{equation}
where $ D$ is  solution of
\begin{equation}
D(R)=-\ln\left\{1-Z\exp\left[-R^2/2-\tilde g D(R)/\pi\right]
\right\}\;.
\label{eq:interact_D}
\end{equation}
Note that the solution $D(R)$ depends only on the fugacity $Z$ and the interaction strength $\tilde g$. The trap frequency and the temperature do not appear explicitly so that the scaling of the atom number with $\omega$ and $T$ (at fixed $Z$) is identical to the result (\ref{eq:ideal_LDA}) for the ideal gas.

Interactions, when treated at the mean-field level, dramatically
change the nature of the solution of eqs.~(\ref{eq:interact_N})-(\ref{eq:interact_D}). For a given trapping frequency $\omega$ and temperature $T$, and for any non-zero $\tilde g$, the atom number $N$ obtained from eq.~(\ref{eq:interact_N}) can be made arbitrarily large by choosing properly the fugacity $Z$. The condensation phenomenon that was obtained in the ideal gas case does not occur anymore. This can be understood qualitatively. For an ideal gas, the saturation of the atom number occurs when the central density in the trap becomes infinite. In presence of repulsive interactions, this singular point cannot be reached and the mean-field treatment provides a solution for any atom number~\cite{Bhaduri:2000}.

\begin{figure}[tbp]
\begin{center}
\includegraphics[width=6cm]{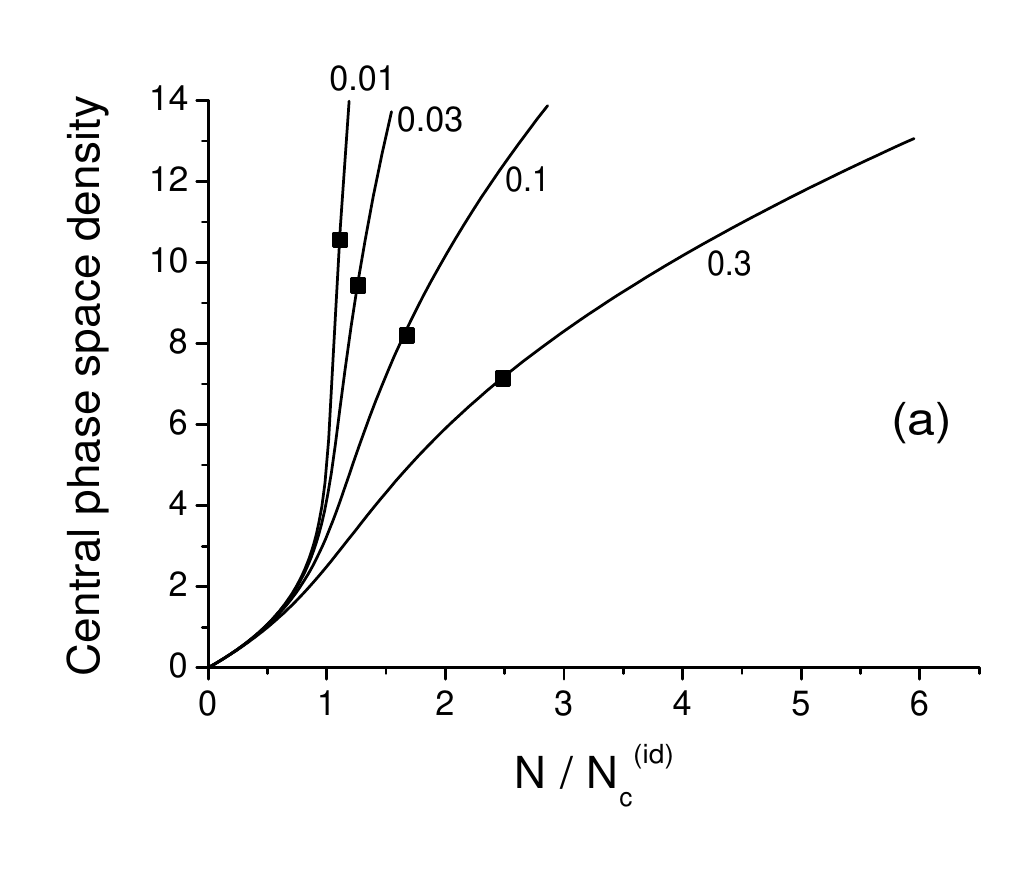}\includegraphics[width=6cm]{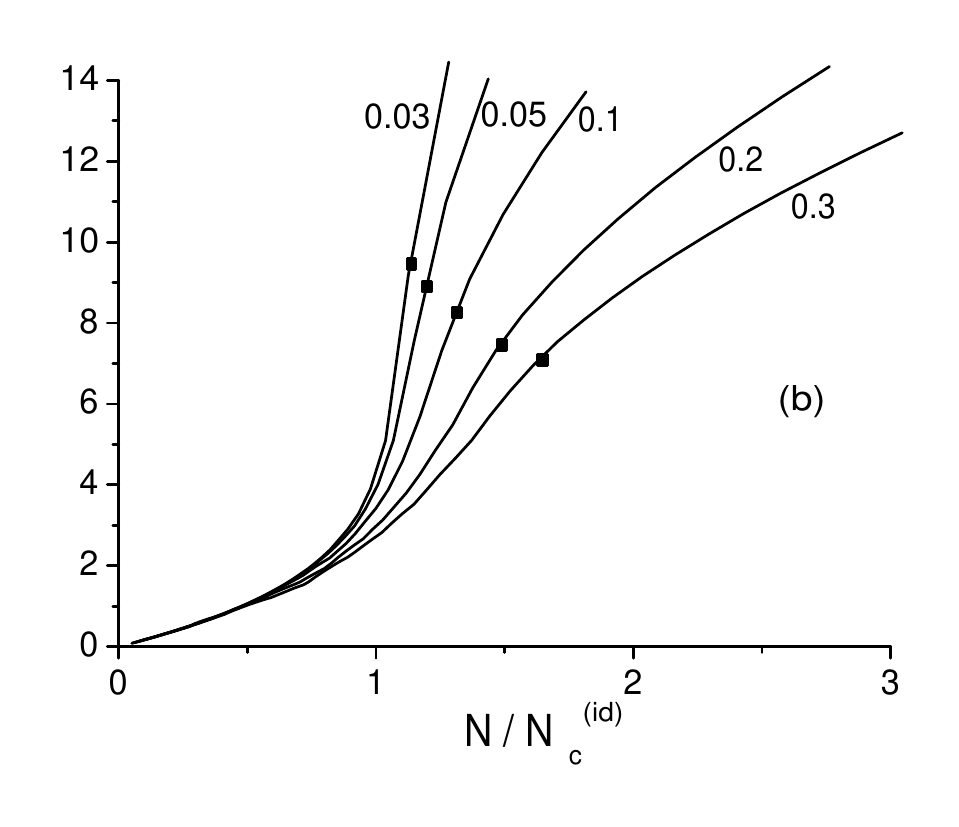}
\caption{Variation of the central phase space density as a function of the atom number, normalized by the critical atom number $\NctwoDid$ for the ideal gas in the same potential and at the same $T$. The value of the interaction strength $\tilde g$ is given for each curve. The black squares indicate the values of $N$ at which the BKT criterion of eq.~(\ref{D_c PS}) is met at the center of the trap. (a) Results obtained using the mean-field Hartree--Fock approach. (b) Results obtained using the bulk results of  \cite{Prokofev:2002} and the local density approximation.}
\label{fig:PSD_Nc}
\end{center}
\end{figure}

We have plotted in figure \ref{fig:PSD_Nc}a the prediction of the mean-field approach for the central phase space density $D(0)$ as a function of the total number of atoms in the trap, for various values of $\tilde g$. As expected, $D(0)$ is a monotonically increasing function of the atom number $N$: more atoms in the trap lead to a larger central density. The other expected feature is that, for a given atom number, $D(0)$ decreases as the repulsion between atoms is increased. In particular the divergence of $D(0)$ that is found in the ideal case for $N=\NctwoDid$ does not show up anymore in presence of repulsive interactions. One could try to push further the mean-field analysis of the equilibrium state, and look for dynamical or thermodynamical instabilities that could appear above some critical atom number~\cite{Ho:1999a,Fernandez:2002,Gies:2004a,Lim:2009}. However we will rather follow the spirit of LDA and assume that the normal to superfluid BKT transition occurs at the center of the trap when the phase space density at this point exceeds the critical value $D_c$ (eq.~(\ref{D_c PS})) predicted for the uniform system \cite{Holzmann:2007a}.

Within the mean-field Hartree--Fock analysis, one can show that, to a very good approximation, the number of atoms that have to be placed in the trap so that the central phase space density reaches the critical value $D_c$ is~\cite{Holzmann:2008}
 \begin{equation}
\frac{\NctwoDmf}{\NctwoDid}=1+\frac{3\tilde g}{\pi^3} \DctwoD^2\;.
 \label{eq:Ncanalytics}
 \end{equation}
An obvious consequence of this result is that for a given trap and a given temperature, the BKT threshold in presence of interactions requires a larger atom number than the BEC of the ideal gas. Equivalently, for a given atom number, the superfluid transition temperature in presence of interaction is lower than the ideal gas condensation temperature. In figure \ref{fig:PSD_Nc}a  we have indicated  with black squares the value of $\NctwoDmf$ for various interaction strengths.

So far we have relied on the mean-field approximation to obtain the relationship between the density $n(r)$ and the local chemical potential
$\mu-V(r)-2gn(r)$. Although this gives a good feeling for the scaling laws that appear in the problem, it cannot provide a very accurate description of the transition. Indeed the mean-field expression $2gn(r)$ for the interaction energy can only be valid at relatively low density, where the density fluctuations are important, so that $\langle n^2 \rangle=2\langle n\rangle^2$. When the density increases and/or the temperature decreases, density fluctuations are reduced and one eventually reaches a situation at very low temperature where those fluctuations are frozen out and $\langle n^2 \rangle=\langle n\rangle^2$. At zero temperature, one expects a quasi-pure condensate in the trap with a density profile given by the Thomas--Fermi law $g n(r)=\mu -V(r)$ (whereas the Hartree--Fock approximation would lead to replacing $g$ by $2g$ in this equation).

To capture the reduction of density fluctuations as the phase space density increases, we now use the numerical results of \cite{Prokofev:2002} obtained using a classical field Monte-Carlo analysis. They provide the value of the phase space density $D$ as a function of $\mu/kT$ in the vicinity of the BKT critical point for a uniform system. Injecting this numerical prediction in the LDA scheme, we obtain the results shown in fig.~\ref{fig:PSD_Nc}b for the central density as a function of the total atom number. As expected this figure is qualitatively similar to the one obtained using the mean-field Hartree--Fock approach. However the classical Monte-Carlo results lead to a noticeable reduction of the critical atom number with respect to the mean field treatment. For example, for $\tilde g=0.15$ (as in the ENS experiment~\cite{Kruger:2007}, see below) the critical atom number for reaching the BKT threshold is expected to be $\sim 1.4 \,\NctwoDid$ using the numerical predictions of~\cite{Prokofev:2002} instead of $\NctwoDmf\sim 1.9 \,\NctwoDid$ using the Hartree--Fock approximation. One might wonder if the classical Monte-Carlo simulations of \cite{Prokofev:2002}, which assume $\tilde g \ll 1$,  remain accurate for the relatively large interaction strength $\tilde g=0.15$. The (positive) answer was given in  \cite{Holzmann:2009b}, which provides a detailed comparison between the predictions of  \cite{Prokofev:2002} and those of a quantum Monte-Carlo simulation of an assembly of trapped bosons for the interaction strength and trapping geometry of the ENS experiment.


\subsection{What comes first: BEC or BKT?}
\label{what comes first in a trap}

In the previous section devoted to the study of a square potential we have explained that the distinction between a conventional BEC transition and a BKT transition is subtle, and we introduced two related concepts:
\begin{itemize}
\item ``BKT-driven condensation", meaning that if the phase space density is increased at constant $\tilde{g}$ the first many-body mechanism encountered is the BKT transition, but due to the resulting slow decay of $g_1$ this transition is accompanied by the appearance of a finite condensed fraction.
\item ``Interaction-enhanced condensation", meaning that there exists a range of phase space densities for which no condensation occurs in an ideal gas but condensation via the BKT mechanism can be induced by increasing the interaction strength from 0 to $\tilde{g}$.
\end{itemize}
In the case of a box potential these two concepts are equivalent. We identified a range of parameters, such that $D_c \ll \ln(4 \pi L^2/\lambda^2)$, for which both effects occur. In the opposite regime of a small system and/or small $\tilde{g}$ neither of the two effects occurs.

The question ``what comes first" is even more subtle in the case of a harmonically trapped gas because of the inhomogeneous density profile. In this case the notions of BKT-driven and interaction-enhanced condensation are not equivalent, and the answer depends on what we keep constant in an experiment, i.e. which path we follow in the phase diagram. Since in a harmonic trap ideal gas condensation occurs even in the thermodynamic limit (i.e. if we neglect the discreteness of single-particle energy levels in the trap), we start by analyzing that case, and separately consider two different experimental paths:

\begin{itemize}
\item
The critical phase space density $D_c$ for a BKT transition (at a fixed non-zero $\tilde{g}$) is finite, while the critical phase space density for ideal gas condensation is infinite. Therefore in a standard experiment where $\tilde{g}$ is kept constant and phase space density is increased, BKT-driven condensation always occurs. In this sense, in practice ``BKT always comes first".

\item
While the critical phase space density for the BKT transition (at non-zero $\tilde{g}$) is lower than for the ideal gas condensation, the critical atom number at fixed $T$ is higher. The critical atom numbers $\NctwoDid$ and $N_{\rm c}^{\rm (BKT)}$ scale similarly with the temperature and the trap frequency ($\propto (k_{\rm B}T/\hbar\omega)^2$), and the ratio $N_{\rm c}^{\rm (BKT)}/\NctwoDid$ is always larger than 1. This can be seen from the mean-field result of eq.~(\ref{eq:Ncanalytics}) or from the Monte-Carlo data shown in figure \ref{fig:PSD_Nc}b. This means that, at fixed $N$, interactions always reduce the transition temperature. Therefore, contrary to the case of a square box potential, we can never have interaction-enhanced condensation.
\end{itemize}

Note that it is not inconsistent that the BKT transition occurs at a lower critical density but higher critical number than the ideal gas BEC, because in a harmonic trap with fixed $N$ and $T$ the peak (phase space) density in a repulsively interacting gas is lower than in an ideal gas. Also note that if we work within the BKT theory and then formally take the $\tilde{g} \rightarrow 0$ limit, we exactly recover the criterion for ideal gas condensation, which is usually derived from a conceptually completely different viewpoint of the saturation of single-particle excited states. We can therefore think of the BEC transition as a special non-interacting limit of the more general BKT theory. This connection naturally emerges when analyzing the case of a harmonically trapped gas, but it could not be made in a uniform system, where the critical temperature for both transitions vanishes in the $\tilde{g} \rightarrow 0$ limit.

Finally, we briefly comment on the case of a realistic experimental harmonic trap, where the spacing of the single-particle energy levels is non-zero. The results for the critical atom numbers for the BKT and the ideal gas BEC transition are essentially unaffected by the non-zero level-spacing. It therefore remains true that interaction-enhanced condensation is not possible. However, the ideal gas BEC in this case occurs at a finite phase space density $D_{\rm BEC}$ in the trap center, which can in principle be lower than $D_c$ for some values of $\tilde{g}$. In this case BKT-driven condensation would also no longer occur, and ``BEC would come first" no matter what path we take in the phase diagram. This scenario is however not relevant for the currently realistic experiments. The value of $D_{\rm BEC}$ is not universal and depends on the details of the trapping potential, but we have evaluated it numerically for a typical trap used in the ENS experiments~\cite{Hadzibabic:2006, Kruger:2007}, and obtained $D_{\rm BEC} \approx 13$~\cite{Hadzibabic:2008}. This means that the condition $D_{\rm BEC} < D_c$ can be fulfilled only in an extremely weakly interacting gas with $\tilde{g} < 10^{-3}$. Experimentally, this regime is essentially indistinguishable from the $\tilde{g} \rightarrow 0$ limit, where the BKT and the BEC transition are no longer distinct.


\subsection{Width of the crossover}
\label{subsec:crossover in the trap}

The divergence of the correlation length that we already discussed in the case of a square box potential (see \S~\ref{width of the crossover in a box}), must also be taken into account.
Suppose that one lowers the temperature of a cold gas in a trap until the BKT threshold density is reached right at the center of the trap. Since the density is everywhere lower than the critical threshold for BKT, LDA would imply that no significant superfluid fraction is present in the gas at this stage. However a significant part of the gas may exhibit a certain degree of coherence as we show now. The critical length (\ref{BKT diverging ell}) is now position dependent, since the critical temperature is a function of the local density. Since $\ell(r)$ is a monotonically decaying function of the distance from the trap center $r$, we can self-consistently assume that the gas is coherent over a region of radius $r_c$ such that $\ell(r_c)=r_c$. We can also provide a crude estimate of $r_c$: $r_c\approx r_T\,(\ln(r_T/\lambda))^{-1/2}$. For practical parameters ($r_T/\lambda\sim 10 \,-\,100$), we find $r_c\sim r_T$, which means that this coherence actually extends over a significant fraction of the cloud when $D=D_c$ at the center of the trap. We can also estimate the width of the cross-over over which the condensed fraction becomes significant. If instead of taking $D=D_c$ at the center of the trap, we take $D=0.7\,D_c$ then $\ell \approx 5\lambda$ at the center of the cloud and no significant coherence exists at this point. The above analysis is confirmed at least qualitatively by numerical simulations performed using a classical field Monte-Carlo analysis. These simulations indeed indicate the emergence of an extended coherence over the cloud at temperatures 10\% to 20\% above the one for which the bulk BKT criterion is met at the trap center~\cite{Bisset:2009b}.


\section{Achieving a quasi-2d gas with cold atoms}
\label{sec:experimental realizations}

The experimental realization of a 2d atomic Bose gas is based on a strongly anisotropic trap with one very tightly confining direction, say $z$, and two more loosely confined degrees of freedom, $x$ and $y$. The $z$ degree of freedom can be considered as frozen from the thermodynamic point of view if the energy gap $\Delta_z$ between the ground state and the first excited state of the $z$ motion is much larger than both $k_{\rm B} T$ and the interaction energy $g n$ (both being typically on the order of one to a few kHz). Since the confinement along $z$ is usually harmonic, with frequency $\omega_z$, the gap is $\Delta_z=\hbar\omega_z$. The $z$ degree of freedom is thermodynamically frozen when the extension of the ground state of the $z$-motion, $a_z=\sqrt{\hbar/m\omega_z}$, is such that $a_z \ll \xi,\lambda/\sqrt{2\pi}$.


\subsection{Experimental implementations}

Conceptually, the simplest scheme to produce a 2d gas is to use a single gaussian light beam that is red-detuned with respect to the atomic resonance. The beam propagates along the $x$ direction, with waists along $y$ and $z$ such that $w_y \gg w_z$, so that it forms a horizontal light sheet. The dipole potential created by this light sheet attracts the atoms towards the focal point, and ensures a strong confinement in the $z$ direction. This technique was used at MIT to produce the first atomic gas (of sodium atoms) in a quasi-2d regime~\cite{Gorlitz:2001b}. More recently it has been implemented at NIST to study the coherence properties of the 2d gas~\cite{Clade:2009}.

One can also produce a 2d gas by using an evanescent light wave at the surface of a glass prism~\cite{Rychtarik:2004,Gillen:2009}, so that the atoms are trapped at a distance of a few micrometers from the horizontal glass surface. The confinement in the horizontal $xy$ plane is provided by an additional laser beam or by a magnetic field gradient. The fact that the confinements in the $xy$ plane and along the $z$ axis have different origins is an interesting feature because it offers the possibility, by releasing only the planar confinement, to study the ballistic expansion of the atoms in the $xy$ plane only (see \S~\ref{subsubsec:twodTOF}).  Another experimental system providing independent confinement in the $xy$ plane and along $z$ has been investigated at Oxford, where a blue detuned, single node, Hermite Gaussian laser beam traps atoms along the $z$ direction, and the confinement in the $xy$ plane is provided by a magnetic field gradient~\cite{Smith:2005}.

Two-dimensional confining potentials that are not based on light beams have also been investigated. One possibility discussed in~\cite{Hinds:1998} consists in trapping paramagnetic atoms just above the surface of a magnetized material that produces an exponentially decaying field. The advantage of this technique lies in the very large achievable frequency $\omega_z$, typically in the MHz range. One drawback is that the optical access in the vicinity of the magnetic material is  not as good as with optically generated trapping potentials. Another appealing technique to produce a single 2d sheet of atoms uses the so-called radio-frequency dressed state potentials~\cite{Zobay:2001,Colombe:2004,Hofferberth:2006}.

A 1d optical lattice setup, formed by the superposition of two running laser waves, is a very convenient way to prepare stacks of 2d gases~\cite{Orzel:2001,Burger:2002,Kohl:2005b,Morsch:2006,Spielman:2007}. The 1d lattice provides a periodic potential along $z$ with an oscillation frequency $\omega_z$ that can easily exceed the typical scale for interaction energy and temperature. The simplest lattice geometry is formed by two counter-propagating laser waves, and provides the largest $\omega_z$ for a given laser intensity. One drawback is that the lattice period is small ($\lambda_0/2$, where $\lambda_0$ is the laser wavelength) so that many planes are generally populated and the addressing of a single plane is difficult. Therefore practical measurements only provide averaged quantities. Another interesting geometry consists in forming a lattice with two beams crossing at an angle $\theta$ smaller than 180$^\circ$~\cite{Hadzibabic:2004}. In this case the distance $\lambda_0/(2\sin(\theta/2))$ between adjacent planes is adjustable, and each plane can be individually addressable if this distance is large enough~\cite{Schrader:2004,Stock:2005}. Furthermore the tunneling matrix element between planes can be made completely negligible, which is important if one wants to achieve a truly 2d geometry and not a periodically modulated 3d system.

Finally, while here we are primarily interested in continuous 2d gases of spinless Bosons, two related experiments on 2d physics also need to be mentioned:

First, an experiment performed in Boulder constitutes a direct implementation of the XY model~\cite{Schweikhard:2007}. There, an array of parallel elongated (quasi-)condensates is created in a 2d optical lattice, and tunneling matrix element $J$ provides a Josephson-type coupling between the neighboring lattice sites. In  this system proliferation of vortices is observed when the temperature is increased. Vortices are detected by turning off the optical lattice and allowing the quasi-condensates trapped on different sites to overlap and interfere. The measured surface density of vortices as a function of the ratio $J/T$ is in good agreement with the BKT theory applied to this system.

Second, in an experiment at Berkeley 2d physics was studied in a spinor Bose--Einstein condensate of Rb atoms with total spin $F=1$ and weak ferromagnetic spin-dependent interactions~\cite{Sadler:2006}. This system is anisotropic, but still 3d with respect to the density degrees of freedom, i.e. the healing length $\xi$ is shorter than the shortest extension of the cloud, along $z$. However, weak spin-dependent interactions correspond to a longer healing length $\xi_s$, so that the system is 2d with respect to the spin degrees of freedom. In this case the magnetization transverse to the quantization axis has a role analogous to the phase of the wave function in a spinless Bose gas. At low $T$ ferromagnetic interactions favor spontaneous symmetry breaking but spin-vortex structures are also observed.


\subsection{Interactions in a 2d atomic gas}
\label{subsection: 2d interactions}

To address the role of interactions in these gases, we start with some considerations concerning the quantum scattering of two atoms when the $z$ motion is strongly confined. In a strictly 2d problem and at low energy, the scattering state between two identical bosonic particles with relative wave vector $\bs k$ is \cite{Adhikari:1986}
\begin{equation}
\psi_{\bs k}(\bs r)\sim e^{i\bs k\cdot\bs r}-\sqrt{\frac{i}{8\pi}}\;f(k)\;\frac{e^{ikr}}{\sqrt{kr}}\ , \qquad
f(k)\approx\frac{4\pi}{-\ln(k^2a_2^2)+i\pi}\;,
\label{eq:scatt_state}
\end{equation}
where $a_2$ is the 2d scattering length. One should notice that contrarily to the 3d case, the scattering amplitude $f(k)$ does not tend to a non-zero finite value when $k$ tends to 0.
In the experimental implementations of 2d gases that have been achieved so far, the confinement along $z$ was still relatively weak from a collisional point of view, in the sense that the thickness $a_z$ of the gas remained notably larger than the 3d scattering length $a_s$. The scattering problem in this confined geometry has been discussed in \cite{Petrov:2000a,Petrov:2001} (see also \cite{Naidon:2007}); the general expression (\ref{eq:scatt_state}) for the scattering state $\psi_{\bs k}$ remains valid and the scattering amplitude can be written
\begin{equation}
f(k)\approx \frac{4\pi}{\sqrt{2\pi}a_z/a_s-\ln(\kappa\,k^2a_z^2)+i\pi}
\label{eq:quasi2d_scatt_amplitude}
\end{equation}
with $\kappa\approx 3.5$, corresponding to the 2d scattering length
\begin{equation}
a_2=a_z\,\sqrt \kappa\, \exp\left(-\sqrt{\frac{\pi}{2}}\,\frac{a_z}{a_s}\right)\ .
\label{eq:2dscattlength}
\end{equation}
Now for all experiments realized so far, the first term $\sqrt{2\pi}a_z/a_s$  in the denominator of eq.~(\ref{eq:quasi2d_scatt_amplitude}) is large compared to 1, and dominates over the logarithmic term $\ln(\kappa \,k^2a_z^2)$ and the imaginary term $i\pi$.  We can then take a constant scattering amplitude (as in 3d) to describe the collisions in the gas: $f(k)\equiv \tilde g \approx \sqrt{8\pi}a_s/a_z$. With this approximation the interaction energy of the gas with density $n(\bs r)$ in the $xy$ plane is
\begin{equation}
E_{\rm int}=\frac{\hbar^2\tilde g}{2m}\int n^2(\bs r)\;d^2r\ .
\label{eq:2d_interation_energy}
\end{equation}
The corresponding values for the 2d scattering length are extremely small, due to the exponential factor in eq.~(\ref{eq:2dscattlength}). Taking for example $a_z=200$~nm and $a_s=5$~nm ($^{87}$Rb atoms), we find $a_2=6\;10^{-29}$~m. Typical surface densities are in the range $10^{13}$~m$^{-2}$, and the dimensionless parameter $na_2^2$ that is relevant for perturbative expansions of the equation of state of the 2d Bose gas~\cite{Schick:1971,Popov:1972,Andersen:2002b,Pricoupenko:2004,Pilati:2005,astrakharchik:2009,Mora:2009} is also extremely small:  $na_2^2\sim 4\;10^{-44}$ for the numbers given above.

The expression (\ref{eq:2d_interation_energy}) can also be obtained by starting from the 3d interaction energy
\begin{equation}
E_{\rm int,3d}=\frac{2\pi\hbar^2a_s}{m}\int n_{\rm 3}^2(\bs r)\;d^3r\ ,
\label{}
\end{equation}
in which we plug directly $n_{\rm 3}(x,y,z)=n(x,y)\exp(-z^2/a_z^2)/\sqrt{\pi a_z^2}$. However the validity condition $a_s \ll a_z$ remains hidden in this procedure.

To summarize,  the collision dynamics in the experiments performed so far is still dominated by 3d physics. The 3d scattering length $a_s$ is much smaller than the thickness of the gas and the scattering amplitude is nearly $k$-independent. This regime is often referred to as `quasi-2d'. It is important to note that the term `quasi-2d' is also used to describe another aspect of the 2d gases: very often the temperature of the gas (and possibly the interaction energy) is not small compared to $\hbar \omega_z$, but comparable or even a bit larger. We discuss in the next subsection how to handle this problem.


\subsection{Residual excitation of the $z$-degree of freedom}

For a quantitative analysis of experiments performed with 2d gases, in particular for the determination of the temperature, it is important to take into account the residual excitation of the $z$-degree of freedom. This was first pointed out in~\cite{Holzmann:2007b}, where a quantum Monte-Carlo simulation gave an estimate for the distortion of the density profile due to this residual excitation. Several possible ways were subsequently proposed to take this excitation into account~\cite{Holzmann:2007b, Holzmann:2008, Hadzibabic:2008, Bisset:2009a,Holzmann:2009b}. The simplest method consists in renormalizing the interaction strength $\tilde g$ to account for the density profile of the gas along the $z$ direction~\cite{Holzmann:2007b}. The predictions derived with this method were compared with Quantum Monte-Carlo results in~\cite{Holzmann:2008} and later analyzed in detail in~\cite{Bisset:2009a}. In the following we outline the slightly more elaborate treatment of~\cite{Hadzibabic:2008} which has the advantage of taking into account not only the thermal excitation of the $z$ degree of freedom, but also the possible deformation of the ground state of the $z$-motion due to atomic interactions.

The method used in~\cite{Hadzibabic:2008} is a direct implementation of the Hartree--Fock approximation (see e.g.~\cite{Kadanoff:1962}) and we first present it for a gas which is uniform in the $xy$ plane. We choose a 3d trial density profile $n_3(z)$ uniform in the $xy$ plane and varying along the strongly confined $z$ direction. We then consider the Hamiltonian with the mean-field energy
\begin{equation}
H=-\frac{\hbar^2}{2m}\nabla^2+\frac{1}{2}m\omega_z^2 z^2 +2\gthreeD n_3(z)\ .
\label{}
\end{equation}
The single particle eigenfunctions of this Hamiltonian can be written $\psi_{\bs k,j}(x,y,z)=\varphi_j(z)\,e^{i(k_x x+k_y y)}\,/\,2\pi$, with energy $E_{\bs k,j}=\hbar^2 k^2/(2m)+\epsilon_j$, where $k^2 = k_x^2 + k_y^2$. The normalized functions $\varphi_j(z)$ and the energies $\epsilon_j$ of the $z$-motion of course depend on the choice of the trial density profile $n_3(z)$. In the Hartree--Fock approximation the average occupation of the single particle level $\psi_{\bs k,j}$ is given by the Bose factor $f(E_{\bs k,j})=(\exp(\beta(E_{\bs k,j}-\mu))-1)^{-1}$. We calculate the corresponding 3d density profile, which is still uniform in $xy$ and has the following $z$ dependence:
\begin{equation}
 n_3'(z)= \sum_j \int d^2k\; |\psi_{\bs k,j}|^2\,f(E_{\bs k,j})=-\frac{1}{\lambda^2}\sum_j |\varphi_j(z)|^2\ln\left(
 1-Ze^{-\beta \epsilon_j}
 \right)\ .
\label{}
\end{equation}
The self-consistency of the Hartree--Fock approximation requires that $n_3(z)$ and $n_3'(z)$ coincide, which can be achieved by iterating the solution of the above set of equations until a fixed point is reached. With this method, we fulfill two goals: (i) We take into account the residual thermal excitation of the levels in the $z$ direction. (ii) Even at zero temperature we take into account the deformation of the $z$ ground state due to interactions. When interactions can be neglected, the eigenstates $\varphi_j(z)$ are the Hermite functions and $\epsilon_j=\hbar\omega_z(j\,+\,1/2)$.

Using the local density approximation, the above method can be straightforwardly adapted to the case where a trapping potential $V_\perp$ is present in the $xy$ plane. The trial density distribution $\nthree({\bs r})$ is now a function of all three spatial coordinates. At any point $(x,y)$, we treat quantum mechanically the $z$ motion and solve the eigenvalue problem for the $z$ variable
 \begin{equation}
\left[\frac{-\hbar^2}{2m}\frac{d^2}{dz^2}+ V_{\rm eff}({\bs r})
\right]\varphi_j(z|x,y)= \epsilon_j(x,y)\;\varphi_j(z|x,y)\ ,
 \end{equation}
where $V_{\rm eff}({\bs r})= V_\perp(x,y)+m \omega_z^2 z^2/2\;
+\;2 \gthreeD n_3({\bs r})$ and $ \int
|\varphi_j(z|x,y)|^2\;dz=1$. Treating semi-classically the $xy$
degrees of freedom, we obtain a new spatial density
  \begin{equation}
\nthree'({\bs r})=-\frac{1}{\lambda^2}\sum_j |\varphi_j(z|x,y)|^2
\ln \left(1-Ze^{-\beta\epsilon_j(x,y)} \right)\ .
  \end{equation}
Again the Hartree--Fock prediction is obtained by iterating this calculation until the spatial density $\nthree({\bs r})$ reaches a fixed point. Since $\varphi_j$ is a normalized function of $z$ at any point $(x,y)$, the total 2d density is
 \begin{equation}
n(x,y) = \int n_3(\bs r)\,dz=-\frac{1}{\lambda^2}\sum_j \ln\left(  1-Ze^{-\beta\epsilon_j(x,y)} \right)\ .
\label{eq:HF3d}
\end{equation}
In the limit where only the ground state $j=0$ of the $z$ motion is populated, the result of this Hartree--Fock approach coincides with the solution of (\ref{eq:interact_2d_trap}). The method used in~\cite{Bisset:2009a} is similar to this approach, but the deformation of the eigenstates due to mean-field interaction was neglected.

The density profiles predicted by this method have been compared with the results of a quantum Monte-Carlo simulation~\cite{Holzmann:2009b}. An important outcome for the analysis of experimental data is the excellent agreement between the two approaches as long as $n\lambda^2<2$. This agreement holds for the temperature regime ($k_{\rm B}T \leq 2 \hbar \omega_z$) and interaction strength ($\tilde g \lesssim 0.15$) relevant for current experiments. The Hartree--Fock approach is therefore well suited for fitting the wings of the experimental density profiles of a quasi-2d gas to extract the temperature and chemical potential.


\section{Probing 2d atomic gases}
\label{sec:experimental probes}

This section is devoted to the presentation of some methods that have been used for the experimental study of 2d Bose gases. We start with the conceptually simplest approach, which consists in the measurement of the steady-state distribution of atoms in a trap. We then turn to the information that can be acquired in a Time-of-Flight expansion. Finally we discuss two schemes that give access to the phase coherence of the gas.


\subsection{\emph{In situ} density distribution}

Conceptually the simplest information that can be obtained on a 2d gas is a picture of the sample along the direction that is strongly confined. Since this degree of freedom is supposed to be frozen out, there is no loss of information due to integration along the line-of-sight. This is in contrast to what happens in 3d, where one has to resort to a non trivial transformation to reconstruct the spatial distribution \cite{Ozeri:2002} (see also \cite{Ketterle:2007} for a review).

Assuming that local density approximation (LDA) is valid, the density distribution in the trap $n(\bs r)$ can be obtained from the equation of state of the homogeneous system. The general form of this equation of state is
 \begin{equation}
 n\lambda^2=F(\mu,k_{\rm B} T,a_2)\ ,
\label{eq:eq_of_state1}
\end{equation}
where $F$ is at this stage an unknown function and $a_2$ is the 2d scattering length. Within LDA the density $n(\bs r)$ in the trap is calculated by replacing $\mu$ by $\mu-V(\bs r)$, where $V(\bs r)$ is the trapping potential.

In the quasi-2d regime that is of practical interest ($a_s\ll a_z$), we have seen in subsection~\ref{subsection: 2d interactions} that the interactions in the gas are characterized to a good approximation by the dimensionless number $\tilde g=\sqrt{8\pi}\,a_s/a_z \ll 1$. In this case eq.~(\ref{eq:eq_of_state1}) can be simplified using dimensional analysis; the expression of the phase space density $D=n\lambda^2$ must take the functional form
 \begin{equation}
 D = G\left( \alpha,\tilde g \right)\qquad \mbox{with}\quad \alpha=\frac{\mu}{\kB T}\ .
\label{eq:eq_of_state2}
\end{equation}
For a gas that is trapped in a harmonic potential $m\omega^2 r^2/2$, the in situ density profile is then given by
 \begin{equation}
n(r) \lambda^2=G\left( \alpha-\frac{r^2}{2r_T^2},\tilde g \right)\ ,
\label{eq:in_trap_density_distrib}
\end{equation}
where we set as above $m\omega^2r_T^2=\kB T$. This expression clearly shows a scale invariance for a given interaction strength $\tilde g$. Suppose that different density profiles $n(r)$ are recorded for various temperatures $T$ and various atom numbers $N$ (hence different chemical potentials $\mu$). According to eq.~(\ref{eq:in_trap_density_distrib}) the profiles can all be superimposed on the same curve $G(\alpha,\tilde g)$, provided they are plotted as a function of $r^2/r_T^2$ and translated along the $x$ axis by the dimensionless quantity $\alpha=\mu/\kB T$. This scale invariance behaviour has been checked with excellent accuracy by M. Holzmann and W. Krauth using quantum Monte-Carlo simulations  \cite{Holzmann:2009}. These simulations were performed for $\tilde g=0.15$, which corresponds to the interacting strength in ENS experiments with Rb atoms.

\begin{figure}[tbp]
\begin{center}
\includegraphics[width=13cm]{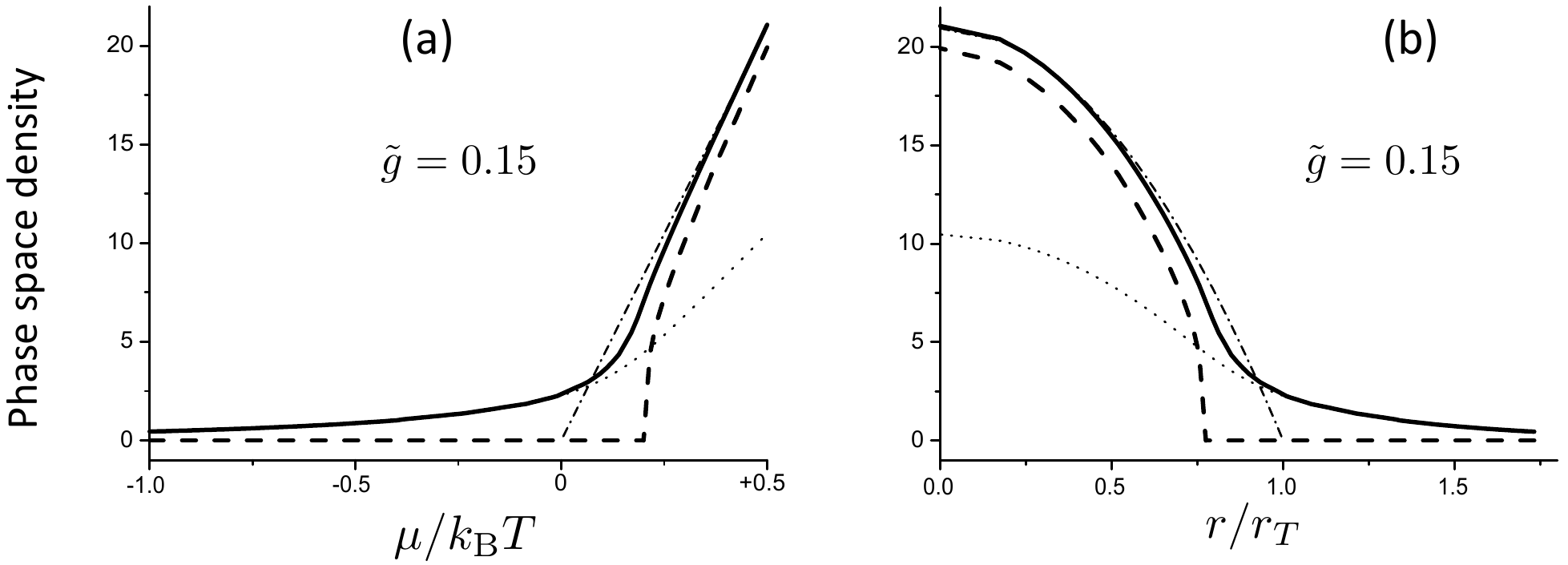}
\caption{(a): Phase space density as a function of $\alpha=\mu/\kB T$ for $\tilde g=0.15$. Continuous line: Total phase space density $D=n\lambda^2$; dashed line: superfluid phase space density $n_s\lambda^2$. The dotted and dash-dotted lines represent the asymptotic regimes for low and high phase space densities, respectively. (b)\emph{In situ} density profiles in a trap  deduced from the left panel using the local density approximation. The plot is made for $\mu/\kB T=0.5$ so that the Thomas-Fermi radius $r_{\rm TF}=\sqrt{2\mu/m\omega^2}$ is equal to $r_T$.
}
\label{fig:eq_of_state}
\end{center}
\end{figure}

For $\tilde g\ll 1$, various asymptotic forms of the function $G(\alpha,\tilde g)$ have been given earlier. When interactions can be neglected ($\tilde g=0$), the equation of state is (cf. eq.~(\ref{no saturation})): $D=-\ln(1-e^\alpha)$. In presence of interactions and for small phase space densities, the mean-field Hartree--Fock method amounts to replace $\mu$ by $\mu-2g n$ into the ideal gas result, which leads to the implicit equation $D=-\ln(1-e^{\alpha-\tilde g D/\pi})$, from which one can extract $D$ as a function of $\alpha$ and $\tilde g$ (see also eq.~(\ref{eq:interact_2d_trap})). In the strongly degenerate limit, where $\mu \gg \kB T$ and $D \gg 1$, density fluctuations are strongly reduced and one expects $\mu = g n$, which can be written as $D=(2\pi/\tilde g) \alpha$. In the intermediate regime, in particular close to the BKT transition point, one can use the results of the classical field Monte-Carlo analysis of \cite{Prokofev:2002}. The resulting function $D=F(\alpha,\tilde g)$ is represented in fig.~\ref{fig:eq_of_state} for $\tilde g=0.15$. The two asymptotic regimes that we just described are indicated with dotted and dash-dotted lines. A remarkable characteristic of the function $F(\alpha, \tilde g)$ is precisely the absence of significant features at the critical point for the BKT transition (corresponding to $\alpha\approx 0.2$ and $D\approx 8$ for $\tilde g=0.15$). This is due to the infinite order of the BKT transition, that does not cause any singularity in the dependance of the total density $n$ on $T$ or $\mu$. On the other hand the superfluid density $n_s$ (plotted as a dashed line in fig.~\ref{fig:eq_of_state}) is discontinuous at the transition point, but this quantity is not directly accessible from an \emph{in situ} measurement. For a detailed comparison between the results of the mean-field Hartree--Fock approach and those obtained from a quantum Monte-Carlo simulation and from a renormalization group treatment, see~\cite{Holzmann:2009b} and~\cite{Lim:2009}, respectively.

Finally we note that the analysis of individual images requires a proper knowledge of the temperature and the chemical potential. These are usually obtained by fitting the wings of the distribution with the appropriate function for the quasi non-degenerate gas (see the discussion following eq.~(\ref{eq:HF3d})). An interesting alternative consists in using in situ density fluctuations to determine these thermodynamic quantities \cite{Zhou:2009}. This promising method that relies on the fluctuation-dissipation theorem for a non-uniform system has not yet been implemented experimentally for a 2d Bose gas.

\subsection{Two-dimensional Time-of-Flight expansion}
\label{subsubsec:twodTOF}

Generally speaking, a Time-of-Flight (TOF) procedure consists in switching off abruptly the potential confining the atoms, letting the cloud expand for an adjustable time, and then measuring the density profile. If the role of interactions is negligible during the expansion, the density profile after a long TOF is proportional to the in-trap momentum distribution. For a two-dimensional system, two types of TOF can be considered. One can switch off the potential confining the atoms in the $xy$ plane, while keeping the strong confinement along the frozen direction $z$; we will call this procedure a ``2d TOF". Alternatively, one can switch off simultaneously the potential in the $xy$ plane and the confinement along $z$, corresponding to a ``3d TOF". We discuss 2d TOF in this section, and 3d TOF in the following one.

We consider here the case of an isotropic harmonic trap in the $xy$ plane $V(r)=m\omega^2r^2/2$. A 2d gas is initially at thermal equilibrium in this trap, with a density profile $n_{\rm eq}(\bs r)$. Suppose that this potential is suddenly switched off at time $t = 0$ whereas the confinement along the $z$ direction remains unchanged. Using the Bogoliubov approach, it was predicted in \cite{Kagan:1996b} that the subsequent evolution of the density distribution is given by the scaling law:
\begin{equation}
n(\bs r,t)=\eta_t^2\, n_{\rm eq}(\eta_t \bs r)\ ,\qquad \eta_t=(1+\omega ^2 t^2)^{-1/2}\ .
\label{eq:expansion_TOF}
\end{equation}
This means that the global form of the spatial distribution is preserved during the TOF. As the expansion proceeds, the interaction energy that was initially present in the gas is converted into kinetic energy in such a way that the density profile at time $t$ is obtained using a scaling transform of the initial one. We emphasize that this remarkable result is stronger than its 3d counterpart \cite{Castin:1996,Kagan:1996b} which holds only in the Thomas-Fermi regime: In the 2d case the scaling behavior is valid both for the superfluid component and for the thermal Bogoliubov excitations. This scaling behavior has been recently observed by the ENS group \cite{Rath:2009}.

The scaling invariance in the expansion of a 2d interacting gas has been explained in terms of the SO(2,1) symmetry group for a whole class of interaction potentials $U(r)$ between two particles \cite{Pitaevskii:1997a}:  it is sufficient that $U$ is a  homogeneous function of degree 2,  $U(\alpha r)=U(r)/\alpha^2$. When this is the case, the result of eq.~(\ref{eq:expansion_TOF}) holds for an arbitrary initial state of the 2d gas,  irrespective of its temperature. The 2d contact interaction potential, which is implicitly assumed in eq.~(\ref{eq:interaction_energy}), belongs to this class of functions.  We note however that a true contact interaction is singular in 2d because it leads to ultraviolet divergences at the level of quantum field equations. A real interatomic potential has a finite range which provides a UV cut-off that eliminates the divergences. This regularization will occur if one uses the more precise treatment of atomic interactions given in eq.~(\ref{eq:quasi2d_scatt_amplitude}). It will lead to deviations with respect to the universal law  (\ref{eq:expansion_TOF}), which remain to be evaluated and characterized.

\subsection{Three-dimensional Time-of-Flight}
\label{subsec:3d tof}
In a 3d TOF both the trapping potential in the $xy$ plane and the strong confinement along the $z$ direction are switched off simultaneously. The physics is then very different from that of a 2d TOF. Along the initially strongly confined direction $z$, the atom cloud expands very fast since the momentum width $\Delta p_z \sim \hbar/\Delta z$ is large. If the atoms are initially in the ground state of a harmonic potential with frequency $\omega_z$ along this axis, the extension of the cloud is multiplied by $\sqrt 2$ in a time $t=\omega_z^{-1}$. The time scale for the expansion of the gas in the $xy$ plane is much longer; it is given by $\omega^{-1}$, where $\omega\ll \omega_z$ is the trapping frequency in this plane. Therefore it is a good approximation to decompose a 3d TOF into two phases. During the first phase, whose duration is a few $\omega_z^{-1}$ (typically 1~ms if $\omega_z/2\pi=3$~kHz), the thickness of the gas along $z$ increases by a factor much larger than 1, but the $xy$ spatial distribution is nearly not modified. At the end of this first phase, the interactions between atoms have become negligible. During the subsequent phase the expansion in the $xy$ plane becomes significant, but on a much longer time scale. It corresponds to the expansion of an ideal gas, whose initial state is equal to the state of the system in the $xy$ plane before the beginning of the TOF.

We now focus on the evolution of the $xy$ degrees of freedom during the second phase, which is essentially governed by single particle physics. The evolution of the density distribution in the $xy$ plane  can be determined from the initial one-body density matrix $g_1(\bs r,\bs r')=\langle \bs r|\rho^{(1)}|\bs r'\rangle$, or from its Fourier transform $\Pi(\bs p)$ with respect to the variable $\bs r-\bs r'$, which represents the momentum distribution in the $xy$ plane.

In absence of any extended coherence in the gas, $g_1(\bs r,\bs r')$ tends to zero when $|\bs r-\bs r'|$ increases, with a characteristic decay length given by the thermal wavelength $\lambda$. The corresponding momentum width is $\Delta p\sim \hbar/\lambda$ and the spatial distribution after TOF will reflect the initial momentum distribution if the TOF duration $t$ is such that $\Delta p\; t/m \gg r_T$, where $r_T$ is the initial size of the gas. For a harmonic confinement in the $xy$ plane, this ``far field" regime corresponds to $\omega t \gg 1$. Taking $\omega/2\pi=30$~Hz as a typical value, the far field regime (say $\omega t>3$) is reached for $t>15$~ms. This corresponds to a typical value for TOF experiments, which thus give access to the momentum distribution in this non (strongly) degenerate regime.

The situation is very different if a significant condensed fraction is present in the gas, as expected in the vicinity and below the BKT transition temperature. In this case we have seen in \S~\ref{subsec:crossover in the trap} that the size $r_c$ of the coherent region of the cloud is $r_c\sim r_T$. The momentum width $\Delta p_c=\hbar/r_c$ of this coherent component is then extremely narrow, and it would require a very long TOF to reach the `far field' regime for this coherent component. Taking $r_c=r_T$ as a typical value, we find that the time $t$ required for a significant expansion of this component, i.e.  $\Delta p_c\;t/m=r_c$, is such that $\omega t=\kB T/\hbar \omega$. For $\omega/2\pi=30$~Hz and $T=100$~nK, this gives $t>300$~ms, which is too long in practice for a TOF.

Therefore in the regime where a relatively strong coherence of the gas is present, a 3d TOF of a realistic duration gives access to a hybrid information. The high energy fraction of the gas is in the far field regime and the wings of the density profile after TOF give access to the large momentum part of the initial state. On the contrary the central feature corresponding to the condensed, superfluid fraction, has not yet undergone a significant expansion. The detailed study of the border between these two components is still a matter of debate. In experiments with rubidium atoms~\cite{Kruger:2007,Cornell:2009}, the density profile after a 3d TOF is well modeled by a two-component distribution and fits with the line of reasoning we just presented. In contrast, in the experiments performed at NIST with a sodium gas~\cite{Clade:2009}, an intermediate third component was introduced in order to obtain a good description of the density profiles after a 3d TOF. This component corresponds to a phase with a spatial scale of coherence that is intermediate between the microscopic length $\lambda$ and the macroscopic one $r_T$, and it is qualified as a ``non-supefluid quasi-condensate" in~\cite{Clade:2009}.

It is interesting to note that 3d TOF is the most common and natural experimental method used in the studies of 3d atomic gases. However, in hindsight, its availability is a non-trivial and rather serendipitous feature of atomic systems for studies of BKT physics. In combination with the finite-size induced condensation, the ability to suddenly turn off the interactions through the fast $z$-expansion provides a much more striking signature of the BKT transition~\cite{Kruger:2007, Clade:2009,Cornell:2009} than one might have theoretically expected. Thinking strictly in 2d, the transition is extremely smooth and one would not naturally expect to see such a dramatic signature in any quantity except the superfluid density. As we discussed in \S~\ref{subsubsec:twodTOF}, in 2d TOF the observed density distribution indeed varies smoothly across the transition.

So far we have discussed the `average' density profile in 3d TOF, which theoretically corresponds to the average of a large number of images obtained under same conditions. It is also interesting to consider the density noise in individual images, which can be related to the phase noise of the gas before expansion. This connection has been exploited for quasi-1d gases since 2001~\cite{Dettmer:2001}. For the 2d case, it has been shown theoretically in~\cite{Imambekov:2009} that the two-point density correlation function after TOF can provide information on the in situ $g_1$ function, at least in the superfluid regime. This method is also specific to 3d TOF, where the phase noise evolves into density noise in interaction-free ballistic expansion.


\subsection{Interference between independent planes}
\label{subsec:interf_two_planes}

\begin{figure}[tbp]
\begin{center}
\includegraphics[width=13cm]{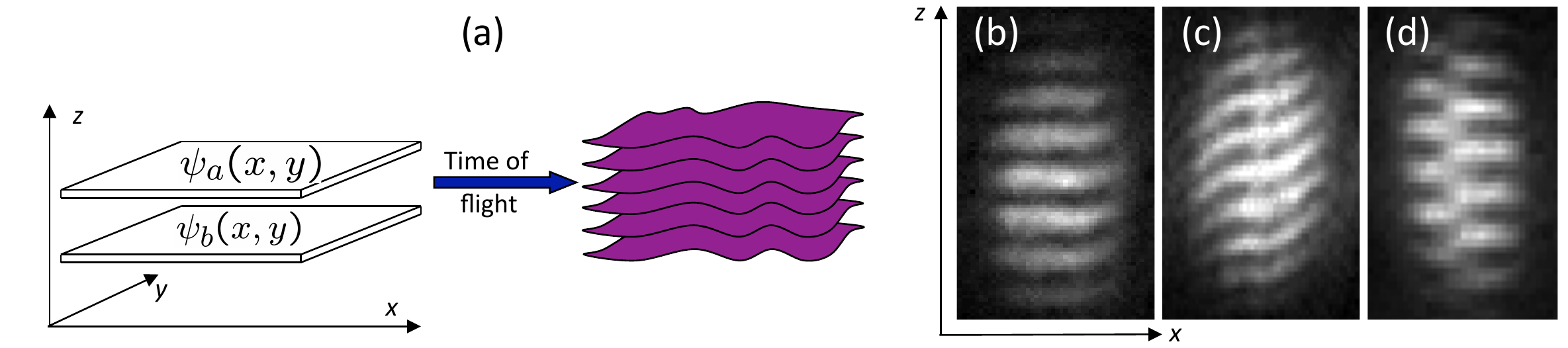}
\caption{(a) Principle of an experiment giving access to the interference between two independent planar gases, observed after time-of-flight. (b-d): Examples of interference patterns measured with the experimental setup described in  \cite{Hadzibabic:2006}. The imaging beam is propagating along the $y$ axis. The pattern (b) is obtained with very cold gases, whereas (c) corresponds to a larger temperature. The dislocation in (d) is the signature for the presence of a vortex in one of the two gases.}
\label{fig:interf}
\end{center}
\end{figure}

Since an important aspect of the physics of 2d Bose gases is related to phase properties, it is natural to investigate measurement schemes based on interferometry. We start with the proposal by Polkovnikov \emph{et al.}~\cite{Polkovnikov:2006a} which showed how a single experimental procedure could characterize both the normal regime (exponential decay of $g_1$) and the superfluid regime (algebraic decay of $g_1$) (see also \cite{Imambekov:2007} for a more complete review). Consider two independent, infinite planar gases located at $z_a=+d_z/2$ and $z_b=-d_z/2$. They are prepared in identical conditions, i.e. they have the same temperature and the same density.  We perform a 3d time-of-flight of duration $t$, that is chosen such that the final extension along $z$ of each cloud is large compared to the initial separation $d_z$ between the planes. The two clouds thus overlap and we want to extract information about the one-body correlation function $g_1$ from their interference pattern (fig. \ref{fig:interf}a).

The state of each plane is described by the wave function $\psi_{a/b}(x,y)$. After expansion, the spatial atomic density $n$ is modulated along any line parallel to the $z$ axis with the period $D_z= ht/md_z$~\cite{Andrews:1997a}:
 \begin{equation}
n \propto |\psi_a|^2 + |\psi_b|^2 + \left(
\psi_a\psi_b^*\, e^{i2\pi z/D_z} \ +\  \mbox{c.c.}
\right)\ .
\label{eq:interf_signal}
\end{equation}
For simplicity we have omitted in the above equation a global envelope factor giving the variation of the density along the $z$ axis. Also we have neglected the expansion in the $xy$ plane during the TOF. As explained above there exists a range of TOF duration where this is valid, if the trapping frequency $\omega$ in this plane is much smaller than $\omega_z$. We see from eq.~(\ref{eq:interf_signal}) that the local (complex) contrast of the density modulation is $\psi_a\psi_b^*$. Experimentally one cannot measure this quantity along a single line, and one rather has access to the average contrast over a region of finite area $A$ in the $xy$ plane. In particular if one performs absorption imaging along the $y$ axis (fig.~\ref{fig:interf}b-d), the image involves an integration of the local contrast $\psi_a\psi_b^*$ along the $y$ direction~\footnote{The length over which the line-of-sight integration occurs can be adjusted by a proper ``slicing" of the cloud just before the imaging process, as in \cite{Andrews:1997a}.}. Averaging the result of this contrast measurement over a large number of realizations, one can define the average contrast ${\cal C}(A)$ :
\begin{equation}
 {\cal C}^2(A)=\frac{1}{A^2}\langle \left| \int_A \psi_a(\bs r)\psi_b^*(\bs r)\;d^2r\right|^2 \rangle\ .
\label{}
\end{equation}
Using the fact that the fluctuations of the wave functions $\psi_a$ and $\psi_b$ are uncorrelated and taking advantage of the translational symmetry of the system,
we find:
\begin{equation}
{\cal C}^2(A)=\frac{1}{A}\int_A |g_1(\bs r)|^2\;d^2 r\ ,
\label{eq:C_square}
\end{equation}
where $g_1(\bs r)=\langle \psi_j^*(\bs r)\psi_j(0)\rangle$ for $j=a,b$. Suppose for simplicity that the area $A$ is a square and consider the two cases of an exponentially decaying $g_1(r) \propto e^{-r/\ell}$, with a characteristic length $\ell \ll \sqrt{A}$ (normal fluid), and an algebraically decaying $g_1(r)\propto r^{-\alpha}$, with an exponent $\alpha<1/4$ (superfluid regime). In the first case, the integral is nearly independent of $A$ and  ${\cal C}^2(A) \propto A^{-1}$. In the second case we find ${\cal C}^2(A) \propto A^{-2\alpha}$ which corresponds to a decay always slower than $A^{-1/2}$. This method is very appealing in the sense that the measurement of a single number, i.e. the exponent $\eta$ characterizing the variation of ${\cal C}\propto A^{-\eta}$, is sufficient to identify the two possible regimes of a 2d Bose gas, and obtain the value of $n_s\lambda^2 = 1/\eta$ in the superfluid case.

A measurement scheme inspired by this method was implemented at ENS \cite{Hadzibabic:2006}, and it indeed revealed a relatively rapid variation of the exponent $\eta$, in qualitative agreement with what is expected near the BKT transition point in the center of the trap. However some notable deviations with respect to the original proposal must be stressed. First, the measurement was performed with anisotropic samples, with lengths $L_y\ll L_x$. The imaging beam was propagating along $y$ and the measured contrast involved a line-of-sight integration over the full length $L_y$~\footnote{The area $A$ was varied by changing the integration distance $\Delta_x$ along $x$. In this case, the BKT transition causes a crossover from $\eta=1/2$ for an exponentially decaying $g_1$ function (with a decay length $\ell\ll \Delta_x$), to  $\eta=1/4$ for a superfluid state.}, which formally breaks the translational invariance that we used to prove eq.~(\ref{eq:C_square}).  Also the presence of a trapping potential in the experiment causes an additional softening of the transition, because of the inhomogeneity of the density along the line-of-sight of the imaging beam. Finally we note that even deep in the superfluid regime where $n_s\lambda^2\gg 1$, the anisotropy of the sample adds some complexity as discussed in \S \ref{subsect:anisotropic_flat_potential}. At large distances ($\Delta_x > n_s\lambda^2\,L_y$), $g_1$ starts to decay exponentially, which complicates the analysis of the dependence of ${\cal C}^2$ on $\Delta_x$. In summary the rapid increase in coherence that occurs in the vicinity of the BKT transition point is sufficiently robust to be revealed experimentally in the average contrast of the interference pattern, but it is difficult to provide a quantitative analysis of the experimental measurements for the variations of ${\cal C}^2(A)$ over a large range of $\Delta_x$.

A subsequent experiment at ENS has compared the conditions for observing a significant interference contrast between the planes and for measuring a clear bimodal density profile after a 3d TOF~\cite{Kruger:2007}. The onsets of the two phenomena were found to coincide within experimental error. Furthermore the spatial part of the gas that gives rise to a visible interference signal coincides with the central, ``non expanding" component of the TOF profile.

An important outcome of the experiments on the interference between two planes is a direct evidence for thermally activated vortices. At low temperatures, long-wavelength phase fluctuations (phonons) result in smooth variations of the phase of the interference fringes, such as seen in fig.~\ref{fig:interf}c. However, if a single isolated vortex is present in one of the two planes while the phase profile of the other plane is smooth, the interference pattern exhibits a sharp dislocation at the coordinate $x$ of the vortex core. Such dislocations have been observed experimentally~\cite{Stock:2005,Hadzibabic:2006,Gillen:2009} and an example is shown in fig.~\ref{fig:interf}d. The occurrence probability of these dislocations has been measured as a function of temperature~\cite{Hadzibabic:2006}. The number of dislocations increases with $T$, until one reaches the temperature at which no interference is visible anymore. Moreover, the relatively sharp increase in the probability of dislocations experimentally coincides with the increase in the exponent $\eta$ characterizing the decay of $g_1$~\cite{Hadzibabic:2006}. Such dislocations also appear in a classical field simulation mimicking the interference between two planar gases~\cite{Simula:2006}. They result from the thermal activation of a vortex pair for which the two members are sufficiently separated from each other. In principle one should also observe in the interference patterns tightly-bound vortex pairs where the two members are separated by $\sim\xi$. However these pairs only shift the fringe pattern by a small fraction of the fringe spacing, which is below the current sensitivity of the experiments.


\subsection{Interfering a single plane with itself}

An interesting alternative to the two-plane interference described above consists in preparing a single plane of atoms and looking at its ``self-interference", using a Ramsey-like method~\cite{Clade:2009}. The gas is initially prepared in an internal state $|1\rangle$. Half of the atoms are coherently transferred into another internal state $|2\rangle$ by a stimulated laser Raman process ($\pi/2$ pulse) that also provides a momentum kick $\bs k_0$ to the atoms. After this process, the part of the cloud in $|1\rangle$ is still globally at rest and the part in $|2\rangle$ moves with the global velocity $\bs v_0=\hbar \bs k_0/m$. After an adjustable time $t$ a second $\pi/2$ Raman pulse remixes the amplitudes of $|1\rangle$ and $|2\rangle$ and provides a momentum kick $\bs k_0 -\bs k_1$. Immediately after this second Raman pulse, one measures the spatial density distribution in $|2\rangle$. This distribution exhibits a modulation along the direction $\bs k_1$, resulting from the interference between the initial state of the cloud and the state displaced by the distance $\bs R=\bs v_0 t$:
\begin{equation}
n(\bs r)\propto |\psi(\bs r)|^2+|\psi(\bs r-\bs R)|^2+ \left(\psi(\bs r)\psi^*(\bs r-\bs R)e^{i\bs k_1\cdot \bs R}+\mbox{c.c.}\right)\ .
\label{eq:self_interf}
\end{equation}
Note that we assume here that no collision occurred during the time $t$ between the part of the cloud at rest in $|1\rangle$ and the part moving at velocity $\bs v_0$ in state $|2\rangle$. This is a valid assumption for the weakly interacting sodium gas of \cite{Clade:2009}.
The modulated density profile in eq.~(\ref{eq:self_interf}) gives a direct access to the function $g_1(\bs r, \bs r-\bs R)$. It can be observed with an imaging beam along the $z$ direction, so that
its measurement does not involve any line-of-sight integration. This method can then reveal finer details than the one presented in \S~\ref{subsec:interf_two_planes}. In particular the authors of
\cite{Clade:2009} could observe a gradual increase of the coherence length $\ell$ of the cloud, as expected from eq.~(\ref{BKT diverging ell}). For small phase space densities the measurement gives $\ell \sim \lambda$, and $\ell$ increases to much larger values when the temperature decreases towards the critical temperature $T_{\rm BKT}$. When $T<T_{\rm BKT}$ a significant interference contrast is observed for all values of $R$ within the size of the central superfluid region.


\section{Conclusions and outlook}
\label{sec:conclusions}

We have reviewed in these notes the theoretical basis for the understanding of the physics of 2d quantum fluids, and discussed some recent experiments performed with atomic gases. These experiments have given access to some aspects of 2d physics that were previously hidden or not measurable in other physical systems, such as the existence of thermally activated individual vortices or the spatial variation of the first order correlation function $g_1(r)$. However a number of issues is still open in the physics of 2d quantum gases, and we outline below some topics which are likely to be of experimental and theoretical interest in the future.

\vskip 3mm \noindent
{\it Higher-order correlation functions.}
The matter-wave interference between two statistically similar, but independent quasi-condensates (such as shown in figure~\ref{fig:interf}), can reveal a wealth of information on the correlations within each individual 2d gas. So far only a fraction of this information has been harnessed, with the study of the average contrast of the interference pattern integrated over some area of interest. A convenient tool for extracting more complete information on $g_1$ as well as higher-order correlation functions is the full statistical distribution of interference contrasts. Two limiting cases can easily be characterized: (i) If the two independent fluids are fully condensed, each image shows a 100\% contrast, with the position of the fringes fluctuating randomly from shot to shot. (ii) If each cloud exhibits only short-ranged correlations, the observed interference results from many uncorrelated fringe patterns along the light of sight, and the distribution of contrasts is an exponential function. For 1d gases, it is possible to describe quantitatively the transition between these two limiting cases~\cite{Gritsev:2006}, and the experimental results~\cite{Hofferberth:2008} are in good agreement with the predictions. In the 2d case, the evolution of the contrast distribution through the BKT transition is still an open problem.

\vskip 3mm \noindent
{\it Out-of-equilibrium dynamical effects.}
Throughout this paper we restricted our discussion to the equilibrium properties of a 2d Bose fluid. The study of dynamical effects, such as transient regimes, can reveal additional information about the system. For example Burkov \emph{et al.}~\cite{Burkov:2007} have studied the dynamics of decoherence between two planar Bose gases, assuming that their local phases are initially locked together, and then the two gases are allowed to evolve independently. This can be achieved experimentally by having a weak potential barrier and hence large tunnel coupling between the two planes for $t<0$, and then suddenly raising the barrier at $t=0$. The contrast of the interference between the two gases gives access to the evolution of the phase distribution under the influence of thermal fluctuations. In~\cite{Burkov:2007} this contrast was shown to decay algebraically at long time, as $t^{-\zeta}$, with the exponent $\zeta$ proportional to the ratio $T/T_{\rm BKT}$. Therefore, in addition to being a stringent test of thermal decoherence in a quantum many-body system, this out-of-equilibrium study could constitute a novel thermometry method. A related phenomenon occurs in 1d systems, where the interference contrast is predicted to decay as $\exp(-(t/t_0)^{2/3})$ (with $t_0$ constant)~\cite{Burkov:2007}, and this prediction is nicely confirmed in the experiments by the Vienna group~\cite{Hofferberth:2007}.

\vskip 3mm \noindent
{\it Transition from 2d to 3d behavior.}
The possibility to vary the tunnel coupling between two or more planar gases can also be used to study the so-called ``deconfinement transition"~\cite{Cazalilla:2007}, corresponding to a gradual evolution from 2d to 3d behavior. The phase coherence between the planes will build up as the strength of the coupling is increased, creating a situation that is reminiscent of the high-$T_{\rm c}$ cuprate superconductors.  For a large number of parallel planes, the deconfinement transition should give rise to a true Bose--Einstein condensate~\cite{Cazalilla:2007}. The two-plane situation is also very interesting, and can lead to the observation of the Kibble-Zurek mechanism~\cite{Mathey:2008}: the superfluid transition temperature is higher for two coupled planes than for a single one, so that sudden switching on of the coupling between the planes (initially in the normal state but close to the single-plane critical temperature) constitutes a quench of the system, and one could observe the subsequent dynamical apparition of a macroscopic quantum (quasi-)coherence.

\vskip 3mm \noindent
{\it Tunable interactions.}
As we have seen throughout the paper, interactions between particles play a crucial role in our understanding of the superfluid phase transition and condensation in 2d fluids. In contrast to the conventional 3d BEC of an atomic gas, where the critical temperature can to a good approximation be predicted using the ideal gas model, the BKT transition is fundamentally interaction-driven. The strength of interactions also affects a variety of other phenomena such as the suppression of density fluctuations in the normal state and the connection between the 2d Bose fluid and the XY model. It would therefore be interesting to revisit the various effects described in these notes while continuously tuning the strength of interactions with a Feshbach resonance~\cite{Tiesinga:1993,Inouye:1998}. In the weak coupling regime ($\tilde{g} < 10^{-1}$) we expect a gradual change from the BKT-dominated to the BEC-dominated behavior, as discussed in Sections~\ref{sec:2d gas in a box} and~\ref{sec:2d gas in a trap}. Further, it would be very interesting to explore the strong-coupling regime ($\tilde{g} > 1$), which is closer to liquid helium films. This regime, which is outside the domain of validity of the Monte-Carlo results \cite{Prokofev:2001,Prokofev:2002}, corresponds to the case where the scattering length $a_s$ becomes comparable the thickness of the sample along the kinematically frozen direction, $a_z$ (see \ref{subsection: 2d interactions}). There the very nature of two-body interactions is expected to change from 3d to 2d~\cite{Petrov:2000a, Petrov:2001,Kestner:2006,Pricoupenko:2008}. Therefore, experimentally reaching the condition $a_s \geq a_z$ would correspond to producing a ``truly 2d" as opposed to a quasi-2d Bose gas.

\vskip 3mm \noindent
{\it Superfluid density.}
Generally speaking, studies of coherence and correlation functions in a 2d fluid, which are well suited to experimental tools of atomic physics, are a natural complement to the ``traditional" studies of superfluidity based on transport measurements, which are well suited to other physical systems such as liquid helium films~\cite{Bishop:1978}.
For example, we have so far assumed that the two types of measurements probe the \emph{same} superfluid density (see e.g. \S~\ref{subsec:interf_two_planes}). However this correspondence may in fact depend on the theoretical model and the exact definition of the superfluid density, and be valid only within the effective low-energy theories. It is therefore important to stress that superfluidity in the traditional transport sense has not yet been directly observed in atomic 2d Bose gases (see e.g.~\cite{Cazalilla:2007}). Establishing atomic 2d gases as experimental systems in which both coherence and transport measurements of superfluidity could be performed would be an important advance, as it would allow experimental scrutiny of the theoretical connections between the two types of probes, and a direct comparison of the different definitions of superfluidity.
Two promising schemes for a direct measurement of the superfluid density (as traditionally defined through transport properties~\cite{Leggett:1973}) in an atomic gas have recently been proposed~\cite{Ho:2009, Cooper:2009}. The first scheme~\cite{Ho:2009} is based on extracting the superfluid density from the in situ density profiles of a rotating 2d gas. The second scheme~\cite{Cooper:2009} is based on using a vector potential generated by Raman laser beams to simulate slow rotation of a gas~\cite{Lin:2009b}, and allows direct spectroscopic measurement of the superfluid density.

\paragraph{Note added in proof} 
A measurement of the equation of state of a 2d Bose gas for various interaction strengths  has just been reported in \cite{Hung:2010}.

\acknowledgments
We warmly thank the directors of the school R. Kaiser and D. Wiersma, as well as the scientific secretary L. Fallani, for organizing this very successful meeting. Many colleagues  helped us with discussions and interactions and the list of those we would like to thank is too long to fit here, but we mention in particular E. Altman, N. Cooper, E. Cornell, E. Demler, B. Dou{\c c}ot, T. Giamarchi, T.-L. Ho, M. Holzmann, M. K\"{o}hl, W. Krauth, W. Phillips, A. Polkovnikov, G. Shlyapnikov, D. Stamper-Kurn, W. Zwerger, as well as the past and present members of the ENS cold atoms group. ZH is supported by EPSRC Grant No. EP/G026823/1. JD is supported by R{\'e}gion Ile de France IFRAF, CNRS, the French Ministry of Research, ANR (Grant ANR-08-BLAN-65 BOFL), and the E.U. project SCALA.  Laboratoire Kastler Brossel is a mixed research unit n$^\circ$ 8552 of CNRS, Ecole normale sup{\'e}rieure, and Universit{\'e} Pierre et Marie Curie.

\bibliographystyle{varenna}

\newpage

\tableofcontents

\end{document}